\documentclass[12pt]{iopart}

\usepackage{graphicx}
\usepackage{subfigure}
\usepackage{verbatim}
\usepackage{dcolumn}
\usepackage{bm}
\usepackage{epsf}
\usepackage{color}
\usepackage[colorlinks=true,citecolor=blue,linkcolor=blue,urlcolor=blue]{hyperref}
\usepackage{hhline}
\usepackage{amssymb}
\usepackage{cite}
\usepackage{tikz}
\usetikzlibrary{arrows,shapes,calc,matrix}

  \expandafter\let\csname equation*\endcsname\relax
  \expandafter\let\csname endequation*\endcsname\relax
\usepackage{amsmath}

\makeatletter
\newcommand\footnoteref[1]{\protected@xdef\@thefnmark{\ref{#1}}\@footnotemark}
\makeatother

\newcommand{\bra}[1]{\left\langle #1\right|}
\newcommand{\ket}[1]{\left|#1\right\rangle}

\newcommand{\bla}{bla\\bla\\bla\\bla\\bla}

\newcommand{\bk}[2]     {\langle #1 | #2 \rangle}
\newcommand{\kb}[2]     {| #1 \rangle \! \langle #2 |}
\newcommand{\cH}        {{\mathcal H}}
\newcommand{\cS}        {{\mathcal S}}
\newcommand{\cA}        {{\mathcal A}}
\newcommand{\cE}        {{\mathcal E}}

\def\be{\begin{equation}}
\def\ee{\end{equation}}
\def\bea{\begin{eqnarray}}          
\def\eea{\end{eqnarray}}
\def\bi{\begin{itemize}}
\def\ei{\end{itemize}}

\usepackage{tikz}
\usetikzlibrary{arrows,shapes}

\newcommand{\draftmode}{1}    
\newcommand{\notetoself}[1]{\ifnum \draftmode=1 {\color[rgb]{0,0,0.8} [#1]} \fi}  
\newcommand{\cuttext}[1]{\ifnum \draftmode=1 {\color[rgb]{0,0.5,0} [#1]} \fi}  
\newcommand{\warntext}[1]{\ifnum \draftmode=1 {\color[rgb]{0.9,0.6,0} #1} \else {#1} \color{black} \fi}

\newcommand\hocom[1]{}

\begin{document}

\tikzstyle{every picture}+=[remember picture]


\title[Eliminating Ensembles from Equilibrium Statistical Physics]{Eliminating Ensembles from Equilibrium Statistical Physics: Maxwell's Demon, Szilard's Engine, and Thermodynamics via Entanglement}

\author{Wojciech H. Zurek}
\address{Theory Division, LANL, MS B213, Los Alamos, NM  87545}

\begin{abstract} 
A system in equilibrium does not evolve -- time independence is its telltale characteristic. However, in Newtonian physics the microstate of an individual system (a point in its phase space) evolves incessantly in accord with its equations of motion. Ensembles were introduced in XIX century to bridge that chasm between continuous motion of phase space points in Newtonian dynamics and stasis of thermodynamics: While states of individual classical systems inevitably evolve, a phase space distribution of such states -- an ensemble -- can be time-independent. I show that entanglement (e.g., with the environment) can yield a time-independent equilibrium in an individual quantum system. This allows one to eliminate ensembles -- an awkward stratagem introduced to reconcile thermodynamics with Newtonian mechanics -- and use an individual system interacting and therefore entangled with its heat bath to represent equilibrium and to elucidate the role of information and measurements in physics. Thus, in our quantum Universe one can practice statistical physics without ensembles -- hence, in a sense, without statistics. The elimination of ensembles uses ideas that led to the recent derivation of Born's rule from the symmetries of entanglement, and I start with a review of that derivation. I then review and discuss difficulties related to the reliance on ensembles and illustrate the need for ensembles with the classical Szilard's engine. A similar quantum engine -- a single system interacting with the thermal heat bath environment -- is enough to 
establish thermodynamics. The role of Maxwell's demon (which in this quantum context resembles Wigner's friend) is also discussed.
\end{abstract}

\maketitle

\section{Introduction}

Thermodynamics is organized logically around equilibrium states -- states in which ``nothing happens''. Yet, in Newtonian dynamics, states of individual systems -- points in phase space -- always evolve, following their trajectories. Therefore, in an individual classical system, incessant motion rules out equilibrium. 

To develop statistical physics -- a bridge between Newtonian dynamics and thermodynamics -- Gibbs \cite{Gibbs} introduced ensembles. 
A classical ensemble is an infinite collection of identical classical systems. It is in equilibrium when states (points in phase space) are distributed so that, while each follows a trajectory in accord with its equations of motion, ``nothing happens'' to the ensemble -- to their phase space distribution. 

This inspired ruse (anticipated to a degree by Boltzmann \cite{Boltzmann} and Maxwell \cite{Maxwell}) was spectacularly successful. It allowed Gibbs to deduce thermodynamics from statistical considerations involving classical ensembles. Yet, it is ultimately unconvincing: An agent dealing with an individual system -- e.g., an engine -- should not be forced to consider a whole collection of systems to deduce properties of his single system. 

By now an extensive mythology legitimizes the ensemble-based approach. Its various rationalizations appeal to ergodicity (the fact that microstates of the system explore rapidly is phase space) or to the ignorance of observers (who do not know specific state of the system, and -- we are told -- are therefore resigned to model it using ensembles). 

These excuses hint at the deficiency of classical physics: It ignores the role of information in defining the state of the system. Information in possession of the observer is subjective -- what the observer knows (the state of his records) is not reflected in the state or in the evolution of the classical system. Maxwell's demon \cite{Maxwell} is in a sense a pre-quantum version of the measurement problem: Considerations involving information are essential to understand why thermodynamics follows from classical dynamics, but the role for information in Newtonian dynamics is limited to the predictive ability of the observer -- the state of the system does not depend on it.

Quantum mechanics does not suffer from this deficiency: Quantum states are {\it epiontic}  \cite{Zurek05} -- 
 they combine epistemic and ontological roles, describing observer's knowledge but also helping determine what exists: What the observer knows is reflected in the state of the system and supplies initial condition for its evolution. However, the key role information plays in quantum physics undermines the solid reality which now gets molded by measurements -- what exists is determined with the help of measurements of what observer decides to find out.

The role of ensembles in statistical physics is to smuggle in probabilities: An ensemble (aka statistical ensemble) is an idealization consisting of a large number of virtual copies (often infinitely many) of a system considered all at once. Each copy represents a possible microstate of the system.  A statistical ensemble is a model of the probability distribution for the microstates of the system. 

There is an obvious kinship between ensembles used in the statistical physics and relative frequency definition of probabilities \cite{vonMises,Gnedenko}. This is no surprise: The need for ensembles arises because thermodynamics uses entropy, which needs probabilities for its definition. In classical settings probabilities and entropy are therefore either subjective, and represent ignorance of an agent, or may have a pretense to be ``objective'' -- e.g., represent frequencies of events in an infinite (hence, hypothetical, and, therefore, ultimately, also subjective!) ``population'' that plays the role of a statistical ensemble.

In quantum physics one can derive Born's rule -- hence, deduce probabilities -- from symmetries of a perfectly known (and, therefore, objective!) state of a system of interest entangled with its environment \cite{Zurek03a,Zurek05,Zurek11,Zurek03b}. This objective definition of probabilities eliminates the need for ``populations'' or subjective ignorance-based excuses: When symmetry implies that every outcome is equally likely, the probability of an ``elementary event''  is $1/N$, where $N$ is the number of such potential outcomes. Counting of equiprobable events replaces counting of members in a population or microstates in an ensemble. The generalization to when events are not equiprobable is straightforward. 

We start, in the next section, with the review of this derivation of Born's rule, as the symmetry of entangled states that allows one to eliminate ensembles is also central to the emergence of probabilities -- of Born's rule -- in our quantum Universe. Given the success of this derivation of Born's rule, it is then natural to inquire whether equilibrium in an individual quantum system can be defined without appeal to ensembles, but, rather, by recognizing symmetries of entanglement with its heat bath. I review this approach \cite{DeffnerZurek2016} with the focus on the demand that equilibrium should represent the absence of motion in an individual system. The main result establishes that any unitary evolution operator acting on a state representing microcanonical equilibrium that results from entanglement with the environment can be ``undone'' in the composite state that represents both the system and the environment by a suitable ``counter-evolution'' in the environment. This leads us to conclude that individual quantum systems entangled with their environment or heat bath provide a natural foundation for thermodynamics in our quantum Universe. We shall also speculate that, in the classical statistical physics, the need to introduce ensembles arose as a reflection of its deficiency -- of its inability to properly recognize the role of information in the fundamental classical dynamics (Newtonian mechanics).

This conclusion is illustrated with Szilard's engine \cite{Szilard}, a paradigmatic classical example of Maxwell's demon operating a one-molecule engine. We also study a similar quantum engine and show how its thermodynamics can be understood without ensembles. 

Szilard's engine was introduced to investigate threats to the second law posed ``by the intervention of intelligent beings'', and we shall start by reviewing some of the discussions it stimulated. The quantum treatment of the molecule in the engine (that, classically, requires an ensemble to represent its various states and make contact with thermodynamics) benefits from the recognition of the role of entanglement. It exposes the thermodynamic consequences of the measurements and removes doubts about the first law of thermodynamics raised by Szilard's engine. 

There is a growing realization that entanglement helps in reaching equilibrium (see e.g.\cite{anders}). 
Thus, the rate of entropy production in a quantum version of a classically chaotic system can be shown to be given by the Kolmogorov-Sinai entropy (i.e., the sum of the positive Lyapunov exponents), and (when the coupling to the environment is weak) it is independent of the details of the system-environment interaction, even though its ultimate origin is the entanglement with the environment\cite{ZP}.  

Some of the discussions of the foundations of statistical physics are based in part on ``typicality'' arguments that employ ensembles of systems entangled with their environments. Using the environment in this fashion one then argues -- relying on randomness, concentration of measure, and, above all, Born's rule involved in tracing out of the environment -- that, {\it typically}, such systems have reduced density matrices close to equilibrium\cite{Popescu,Goldstein}.  

These studies appeal to ensembles (an artifice that we aim to do without). Moreover, they conflate questions about the approach to equilibrium (an issue we return to in Section 6, as it is clearly of interest \cite{ZP}) and its static nature. However, absence of evolution in equilibrium is our present focus), and we shall justify it without appeal to ensembles -- a unique state should suffice to represent equilibrium in a single, unique system. By contrast, {\it ``typicality'' relies on ensembles to define what is typical}. It starts with different assumptions than our approach and answers a different (but not unrelated) set of questions.

We briefly comment on these approaches later. For now we note that our goal is to show, with minimal ingredients (i.e., without decoherence, randomness, concentration of measure, typicality, or even Born's rule) how entanglement can eliminate evolution in a {\it single system}.  Thus, to define equilibrium one does not need ensembles: The symmetry of entanglement suffices.

\subsection{Core Quantum Postulates}

It is best to start from well-defined solid ground. To this end, we have extracted the
list of quantum postulates that are explicit in Dirac \cite{Dirac}, and at least implicit in many quantum textbooks.

The first two postulates are ``purely quantum":

(i) {\it The state of a quantum system $\cS$ is represented by a vector in its Hilbert space} $\cH_{\cS}$.

(ii) {\it Evolutions are unitary (e.g., generated by the Schr\"odinger equation).}

These postulates provide an essentially complete summary of the mathematical structure of the theory.
They are sometimes supplemented by a composition postulate: 

(o) {\it States of composite quantum systems are represented by a vector in the tensor product of the Hilbert spaces of its components.}

We cite it here
for completeness, and note that physicists differ in assessing how much of postulate (o) follows from (i).
We shall not be distracted by this minor issue and move on to where the real problems are.  Readers
can follow their personal taste in supplementing (i) and (ii) with whatever portion of (o) they deem
necessary.

Using (o), (i) and (ii), and suitable Hamiltonians, one can  calculate everything that can be calculated in quantum theory. Yet, such calculations would only be a mathematical exercise 
-- one can predict nothing of experimental consequence from their results. What is missing is a connection with physics -- a connection with measurements. 

A way to establish such correspondence 
between abstract state vectors in $\cH_{\cS}$ and laboratory experiments 
(and/or everyday experience) is needed to relate quantum mathematics to physics. The task of establishing this correspondence starts with the next postulate:

(iii) {\it Immediate repetition of a measurement yields the same outcome.}

Postulate (iii) is idealized (it is hard to devise such non-demolition measurements, but in principle it
can be done). Yet it is essential and uncontroversial. The very notion of a ``state'' is based on predictability, i.e., something like the repeatability postulate (iii): the most rudimentary prediction is 
a confirmation that the state is what it is known to be. Moreover, a classical equivalent of 
(iii) is taken for granted (even unknown classical states can be discovered without getting re-prepared), so there is no clash with our classical intuition here.

One more important role played by the postulate (iii) is to supply a special ``normative'' case of probability: Repeatability implies that one is {\it certain} to obtain the same outcome. Thus, one now has smuggled in -- thanks to postulate (iii) -- probability in its most rudimentary instantiation, when ``$p=1$''.

Postulate (iii) ends the uncontroversial part of the list of postulates. This collection of postulates comprises our {\it quantum core} -- our {\it credo}, the set of beliefs that will be the foundation of our quantum theory of the classical. 

The other two postulates usually listed in textbooks -- the collapse postulate (iv) and the Born's rule (v) -- are rightly regarded as controversial. We list them here for completeness, but we shall not use them as assumptions. 

(iv) {\it Measurement outcomes are limited to an orthonormal set of states (eigenstates of the measured observable). In any given run of a measurement an outcome is just one such state.}

(v) {\it The probability $p_k$ of an outcome $\ket {s_k}$ in a measurement of a quantum system that was
previously prepared in the state $\ket \psi$ is given by $|\bk {s_k} \psi |^2$}.

This last postulate in our list fits very well with Bohr's ``Copenhagen Interpretation'' approach to the quantum - classical transition, and, especially,
with postulate (iv) \cite{Bohr}.  However -- as was noted by e.g. Everett \cite{Everett} -- it is at odds with the spirit of the relative state
approach, or any approach that attempts (as we do) to deduce our perception of the classical everyday reality starting from the quantum laws that govern our Universe. 

This does not mean that there is a mathematical inconsistency in the Copenhagen Interpretation -- one can certainly
use Born's rule (as the formula $p_k= |\bk {s_k} \psi |^2$ is known) along with the relative state
approach in averaging to get expectation values and thus obtain the reduced density matrix. Indeed, decoherence was at least initially practiced \cite{Zurek81,Zurek91,GJK+96a} without worrying about the origin of Born's rule. Everett's point was,
rather, that Born's rule should be {\it derivable} from the other axioms of quantum theory, and we shall now show how to do that. 

We also note that many (but not all) of the consequences of the collapse postulate (iv) can be deduced from the unitary evolution \cite{Zurek14}. Thus, while unitarity is inconsistent with ``literal collapse'' it can (along with the other core quantum postulates) account for the discreteness that underlies the perception of quantum jumps \cite{Zurek07,Zurek13}.

\section{Probabilities from Entanglement}

Several past attempts at the derivation of Born's rule turned out
to be circular. Here we present the key ideas behind a circularity-free approach. Thus, we briefly
recount some salient points of a recent derivation of Born's rule based on a symmetry of
entangled states -- on {\it entanglement - assisted invariance} or {\it envariance}.
The study of envariance
as a physical basis of Born's rule started with \cite{Zurek03a,Zurek05,Zurek11,Zurek03b}, and is now the focus of
several other papers (see e.g. Refs. \cite{53,5,H,Zurek14}). The key idea is illustrated in Fig. 1 and Fig. 2, and there are now several experiments \cite{Laflamme,Ebrahim,Deffner,Ferrari} that test some of the tenets of the derivation we are about to present.

Envariance also accounts for the loss of the physical significance of local phases between Schmidt
states (in essence, for decoherence)\footnote{Schmidt decomposition expresses a pure state of the composite bipartite system in terms of the orthogonal basis states of the two subsystems. It involves only as many states as there are in the smaller of the two subsystems. Thus, a general pure state of $\cS$ and $\cE$ can be always written as a sum 
$$ | \Psi_{\cal SE}\rangle \propto \sum_{k=1}^K \alpha_k
|s_k\rangle |\varepsilon_k\rangle \ $$
where $\alpha_k$ are complex and Schmidt states of the system and of the environment are orthogonal. The number of states in such a sum is no larger than the dimension of the smaller of the two Hilbert spaces $\cH_\cS$ and $\cH_\cE$ and is therefore much smaller than the dimension of $\cH_{\cS\cE}$. Moreover, spectra of the density matrices of $\cS$ and $\cE$ are identical and given by $|\alpha_k|^2$. 
Any pure state of two systems can be always written as a Schmidt decomposition. 
We shall rely on this below.}. The eventual loss of coherence between pointer states
can therefore be regarded as a consequence of quantum symmetries of the states of systems entangled
with their environment. So, the essence of decoherence arises from symmetries of entangled states,
and certain aspects of {\it environment-induced superselection} \cite{Zurek82} or {\it einselection}
can be studied without
employing the usual tools of decoherence theory (reduced density matrices and trace) that, for their physical significance, rely on Born's rule \cite{Zurek14}.

Envariance also allows one to justify additivity of probabilities, while the derivation of Born's rule 
by Gleason \cite{Gleason} assumed it (along with the other of Kolmogorov's axioms of the measure-theoretic formulation of the foundations of probability theory, and with the Copenhagen-like setting). 
By contrast, appeal to symmetries allows one to deduce additivity, also in the classical setting (as was done by Laplace: see \cite{Gnedenko}). Moreover,
Gleason's theorem (with its rather complicated proof based on ``frame functions'' introduced especially for this purpose) provides no motivation as to why the measure one obtains should have any
physical significance -- i.e., why should it be regarded as probability? As illustrated in Fig. 2 and discussed below,
the envariant derivation of Born's rule has a transparent physical motivation.

The additivity of probabilities is a highly nontrivial point. In quantum theory the overarching
additivity principle is the quantum principle of superposition. Anyone familiar with the double slit
experiment knows that probabilities of quantum states (such as the states corresponding to passing
through one of the two slits) do {\it not} add, which in turn leads to interference patterns. 

The presence of entanglement eliminates
local phases (thus suppressing quantum superpositions, that is, doing the job of decoherence). This leads to additivity of probabilities of events associated with preferred pointer states. 

\subsection{Decoherence, phases, and entanglement}

Decoherence is the loss of phase coherence between preferred states. It occurs when $\cS$ starts in a superposition of pointer states singled out by the interaction (represented below by the Hamiltonian ${{\bf H}_{\cS\cE}}$). The states of the system leave imprints -- become `copied'  by $\cE$, its environment:
$$
(\alpha \ket \uparrow + \beta \ket \downarrow)\ket {\varepsilon_0} { \stackrel {{\bf H}_{\cS\cE}} \Longrightarrow } \alpha \ket \uparrow \ket {\varepsilon_\uparrow}  + \beta \ket \downarrow \ket {\varepsilon_\downarrow} = \ket {\psi_{\cS\cE}}. \eqno(1)
$$
One can show (see \cite{Zurek07,Zurek13}) that the  states untouched by such copying must be orthogonal, $\bk \uparrow  \downarrow = 0$. Their 
superposition,
$\alpha \ket \uparrow + \beta \ket  \downarrow$ 
turns into an entangled $\ket {\psi_{\cS\cE}}$. Thus, neither $\cS$ nor $\cE$ alone has a pure state. This loss of purity signifies decoherence. One can still assign a mixed state that represents the surviving information about $\cS$ to the system. 

In a superposition phases matter: In a spin $\frac 1 2$--like state $\cS$ $\ket \rightarrow = \frac {\ket \uparrow + \ket \downarrow} {\sqrt 2}$ is orthogonal to $\ket \leftarrow =\frac { \ket \uparrow - \ket \downarrow} {\sqrt 2}$.  The phase shift operator ${\bf u}^{\varphi}_\cS=\kb \uparrow \uparrow +  e^{\i \varphi}  \kb \downarrow \downarrow$ alters the phase that distinguishes them: for instance, when $\varphi=\pi$, it converts $\ket  \rightarrow $ to $\ket \leftarrow$. In experiments ${\bf u}_\cS^\varphi$ would shift the interference pattern.

We assumed perfect decoherence, $\bk {\varepsilon_\uparrow}{\varepsilon_\downarrow} = 0$: $\cE$ has a perfect record of pointer states.
Phase changes can be detected. 
What information survives decoherence, and what is lost?

Consider someone who knows the initial 
pre-decoherence state, $\alpha \ket \uparrow + \beta \ket \downarrow$, and would like to make predictions about the decohered $\cS$. 
We now show that 
when $\bk {\varepsilon_\uparrow}{\varepsilon_\downarrow} = 0$ 
the phases of $\alpha$ and $\beta$ no longer matter for $\cS$ -- $\varphi$ has no effect on {\it local} state of $\cS$, so measurements on $\cS$ cannot detect a phase shift, as there is no interference pattern to shift.

The phase shift 
${\bf u}_\cS^\varphi \otimes {\bf 1}_\cE$ (acting on an entangled $\ket {\psi_{\cS\cE}}$) 
cannot have any effect on its local state because it
can be undone by ${\bf u}_{\cE}^{-\varphi}=\kb {\varepsilon_\uparrow} {\varepsilon_\uparrow} + e^{-\i \varphi} \kb {\varepsilon_\downarrow}{\varepsilon_\downarrow}$, a `countershift' acting on a distant $\cE$ decoupled from the system: 
$${\bf u}_{\cE}^{-\varphi}({\bf u}_\cS^{\varphi} \ket {\psi_{\cS\cE}})={\bf u}_{\cE}^{-\varphi}(\alpha \ket \uparrow \ket {\varepsilon_\uparrow}  + e^{\i \varphi}\beta \ket \downarrow \ket {\varepsilon_\downarrow})=\ket {\psi_{\cS\cE}} . \eqno(2) $$ 
As the phases in $\ket {\psi_{\cS\cE}}$ can be changed in a faraway $\cE$
decoupled from but entangled with $\cS$, 
 they can no longer influence local state of $\cS$. (This follows from quantum theory alone, but is essential for causality -- if they could, measuring $\cS$ would reveal this, enabling superluminal communication!)


The loss of phase coherence is decoherence. Superpositions 
decohere as $\ket \uparrow, \ket \downarrow$ are recorded by $\cE$. 
This is not because phases become ``randomized'' by interactions with $\cE$, as is sometimes said. Rather, 
they become delocalized: they lose significance for $\cS$ alone. They 
no longer belong to $\cS$, so measurements on $\cS$ cannot distinguish states that started as superpositions with different phases for $\alpha, \beta$. 
Consequently, information about $\cS$ is lost -- it is displaced into correlations between $\cS$ and $\cE$, and local phases of $\cS$ become a global property -- global phases of the composite entangled state of $\cS\cE$. 

We have considered this information loss here without reduced density matrices, the usual decoherence tool. Our view of decoherence appeals to symmetry, invariance of $\cS$ -- {\it en}tanglement-assisted in{\it variance} or {\it envariance} under phase shifts of pointer state coefficients, Eq. (2). As $\cS$ entangles with $\cE$, its local state becomes invariant under transformations that could have affected it before. 


The rigorous proof of coherence loss uses quantum core postulates (o)-(iii) and relies on {\bf facts 1 -- 3}. These core postulates
imply:

\vspace{1.5mm}

{\bf 1. A unitary must act on a system to change its state.} The state of $\cS$ that is not acted upon doesn't change even as other systems evolve (so ${\mathbf 1}_\cS \otimes (\kb {\varepsilon_\uparrow} {\varepsilon_\uparrow} + e^{-\i \varphi} \kb {\varepsilon_\downarrow}{\varepsilon_\downarrow})$
does not affect $\cS$ even when ${\cS\cE}$ are entangled, in $\ket {\psi_{\cS\cE}}$);
\vspace{1.5mm}

{\bf 2. The state of a system is all there is to predict measurement outcomes}; 

\vspace{1.5mm}

{\bf 3. A composite state of the whole determines states of subsystems} (so the local state of $\cS$ is restored when the state of the whole $\cS\cE$ is restored). 

\vspace{1.5mm}

The {\bf facts} help characterize local states of entangled systems without using reduced density matrices. 
Thus, the phase shift ${\bf u}_\cS^\varphi \otimes {\mathbf 1}_\cE=(\kb \uparrow \uparrow +  e^{\i \varphi}  \kb \downarrow \downarrow)  \otimes {\mathbf 1}_\cE$ 
acting on pure pre-decoherence state matters -- 
measurement can reveal $\varphi$. In accord with facts 1 and 2, 
${\bf u}_\cS^\varphi$ changes $\alpha \ket \uparrow+ \beta \ket \downarrow$ into $\alpha \ket \uparrow+e^{\i \varphi}  \beta \ket \downarrow$.
However, the same ${\bf u}_\cS^\varphi$ acting on $\cS$ in an entangled state $\ket {\psi_{\cS\cE}}$ does not matter for $\cS$ alone, as it can be undone by ${\mathbf 1}_\cS \otimes (\kb {\varepsilon_\uparrow} {\varepsilon_\uparrow} + e^{-\i \varphi} \kb {\varepsilon_\downarrow}{\varepsilon_\downarrow})$, a countershift 
acting on a faraway, decoupled $\cE$. As the global $\ket {\psi_{\cS\cE}}$ is restored, by fact 3 the local state of $\cS$ is also restored even if $\cS$ is not acted upon (so by fact 1, it remains unchanged). 
Hence, the local state of decohered $\cS$ that obtains from $\ket {\psi_{\cS\cE}}$ 
could not have changed to begin with, and so it cannot depend on phases of $\alpha, \beta$.

The only pure states invariant under such phase shifts (unaffected by decoherence) are pointer states \cite{Zurek81}.
This resilience 
lets them preserve correlations.
For instance, entangled state of the measured $\cS$ and the apparatus 
decoheres as $\cA$ interacts with $\cE$:
$$(\alpha \ket \uparrow \ket {A_\uparrow}  + \beta \ket \downarrow \ket {A_\downarrow} ) \ket {\varepsilon_0}
{ \stackrel {{\bf H}_{\cA\cE}} \Longrightarrow }
\alpha \ket \uparrow \ket {A_\uparrow} \ket {\varepsilon_\uparrow}  + \beta \ket \downarrow \ket {A_\downarrow} \ket {\varepsilon_\downarrow} = \ket {\Psi_{\cS\cA\cE}} $$
The pointer states $\ket {A_\uparrow}, \ket {A_\downarrow}$ of $\cA$ survive decoherence by $\cE$. They retain perfect correlation with $\cS$ (or an observer, or other systems) in spite of $\cE$, independent of the value of $\bk {\varepsilon_\uparrow} {\varepsilon_\downarrow}$. Stability under decoherence is -- in our quantum Universe -- a prerequisite for effective classicality: Familiar states of macroscopic objects also have to survive monitoring by $\cE$ and, hence, retain correlations.

The decohered $\cS\cA$ is described by a {\it reduced density matrix}, 
$$ \rho_{\cS\cA} = \Tr_\cE \kb {\Psi_{\cS\cA\cE}}{\Psi_{\cS\cA\cE}} \ . $$ 
When $\bk {\varepsilon_\uparrow} {\varepsilon_\downarrow} = 0$, the pointer states of $\cA$
retain correlations with the outcomes:
$$ \rho_{\cS\cA} 
= |\alpha|^2 \kb \uparrow \uparrow \kb {A_\uparrow} {A_\uparrow} + |\beta|^2 \kb \downarrow \downarrow \kb {A_\downarrow}{A_\downarrow}  $$
Both $\uparrow$ and $\downarrow$ are present: There is no `literal collapse'.
We will use $\rho_{\cS\cA}$ to examine information flows. Thus, we will need the probabilities of the outcomes.

Taking the trace is a mathematical operation. However, regarding reduced density matrix $\rho_{\cS\cA}$ as statistical mixture of its eigenstates -- states $\uparrow$ and $\downarrow$ and $A_\uparrow, A_\downarrow$ (pointer state) records -- relies on Born's rule, that lets one view tracing as averaging. 
We didn't use it until just above to avoid circularity. Now we derive $p_k=|\psi_k|^2$, Born's rule: We need to prove that the probabilities are indeed given by the eigenvalues $|\alpha|^2,  |\beta|^2$ of $\rho_{\cS\cA}$.  
This is the postulate (v), obviously crucial for relating our quantum formalism to experiments. We want to deduce Born's rule from the quantum core postulates (o)-(iii). 

\subsection{Probabilities from symmetries of entanglement}

\begin{figure*}[tb]
\begin{tabular}{l}
\vspace{-0.15in}
\includegraphics[width=5.8in]{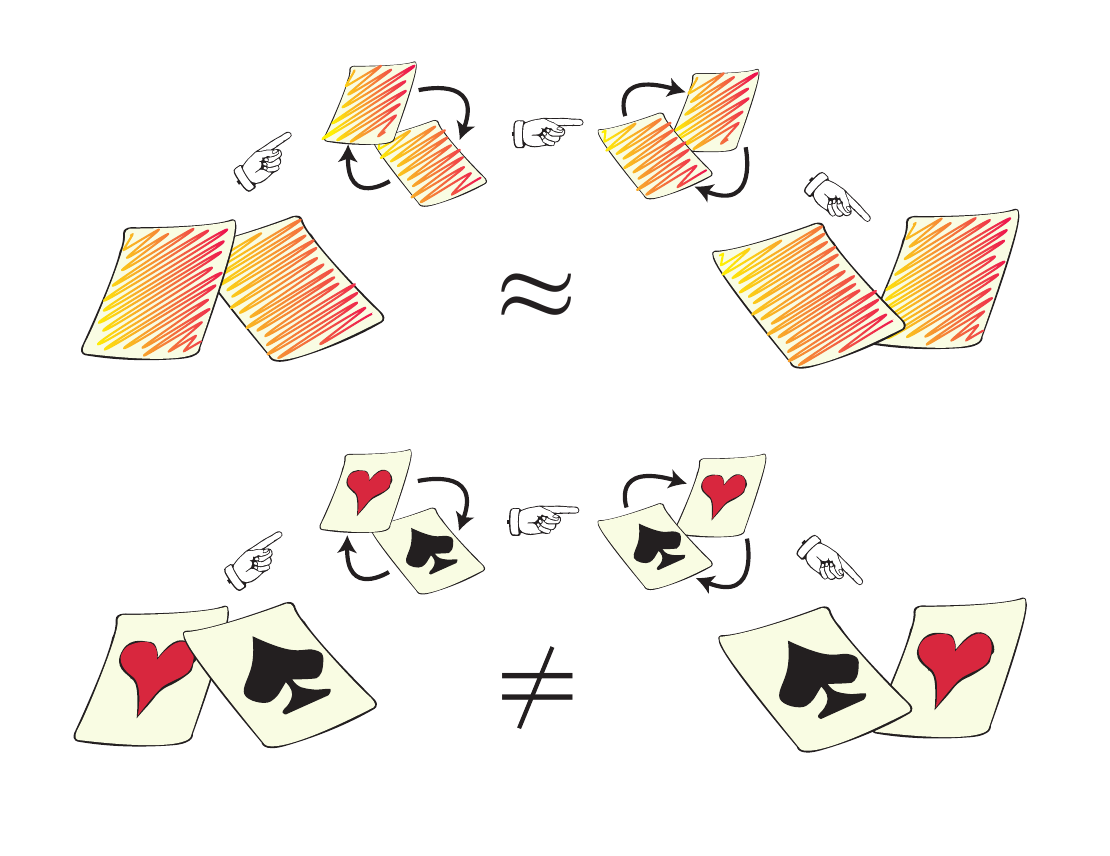}\\
\end{tabular}
\caption{Envariance is a symmetry of entangled states. It allows one to demonstrate Born's rule \cite{Zurek03a,Zurek05,Zurek11,Zurek03b} using a combination of (i) an old intuition of Laplace \cite{40}
about invariance and the origins of probability and (ii) quantum symmetries of entanglement.
{\bf (a)} Laplace's {\it principle of indifference} (illustrated with playing cards) aims to establish
 symmetry using invariance under swaps.
A player who doesn't know face values of cards is  indifferent -- does not care -- if they are swapped before
he gets the one on the left. For Laplace, this indifference was the evidence of invariance, hence, of a (subjective) symmetry: 
It implied {\it equal likelihood} -- equal probabilities of the invariantly swappable
alternatives. For the two cards above, subjective probability $p_\spadesuit ={ \frac 1 2}$ would be inferred by someone who does not know their face value, but knows that one and
only one of the two cards is a spade. 
When probabilities of a set of elementary events are provably equal, one
can compute probabilities of composite events and thus develop a theory of probability. Even 
additivity of probabilities can be {\it established} (see, e.g., Gnedenko, \cite{Gnedenko}).
This is in contrast to Kolmogorov's measure-theoretic
axioms (which {\it include} additivity of probabilities). Above all, Kolmogorov's theory does not assign probabilities to elementary events (physical or otherwise), while our approach yields probabilities when symmetries of elementary events under swaps are known (see Fig. 2).
{\bf (b)} The problem with Laplace's indifference is its subjectivity:
The actual physical state of the system (the two cards) is altered by
a swap.  A related problem is that the assessment of indifference is based on ignorance: It
as was argued, e.g., by supporters of the relative frequency approach (regarded by many as more ``objective'' foundation of probability) that it is impossible to deduce anything 
(including probabilities) from ignorance. This is (along with subjectivity) was a reason why equal likelihood
was regarded with suspicion as a basis of probability in classical physics.}
\label{ccards}
\end{figure*}

\begin{figure*}[tb]
\begin{tabular}{l}
\includegraphics[width=6.5in]{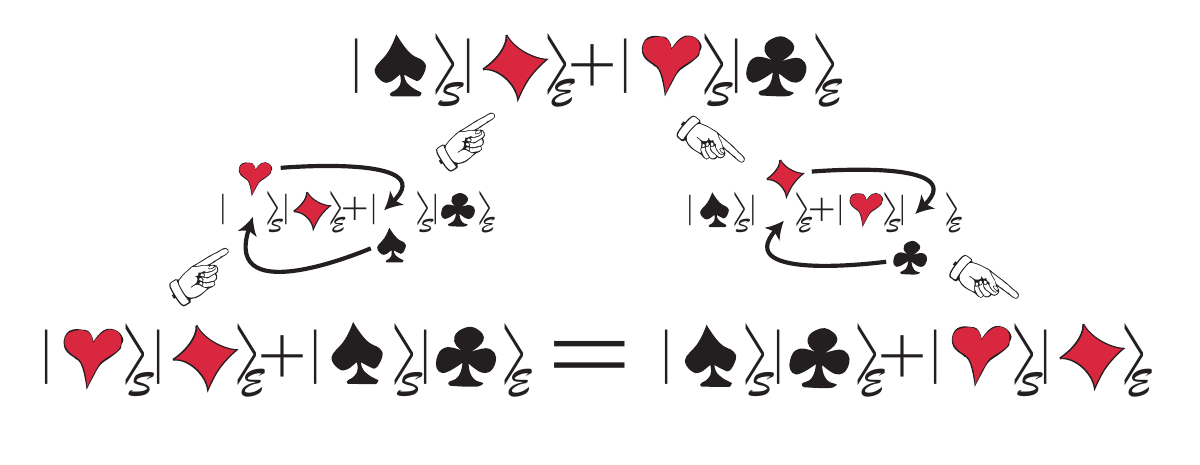}\\
\end{tabular}
\caption{In quantum physics symmetries of entanglement can be used to deduce objective
probabilities starting with a known state. The relevant symmetry is the {\it en}tanglement - assisted in{\it variance} or {\it envariance}.
When a pure entangled state of a system $\cS$ and another system we call ``an environment $\cE$'' (anticipating connections with decoherence) $
|\psi_{\cal SE}\rangle = \sum_{k=1}^N a_k |s_k\rangle |\varepsilon_k\rangle $ can be transformed
by $U_{\cal S}=u_{\cal S} \otimes {\bf 1}_{\cal E}$ acting solely on ${\cal S}$, but the effect of $U_{\cal S}$ can be undone by acting solely on ${\cal E}$ with an appropriately chosen $U_{\cal E}=
{\bf 1}_{\cal S} \otimes u_{\cal E}$, $U_{\cal E}|\eta_{\cal SE}\rangle  = ({\bf 1}_{\cal S} \otimes
u_{\cal E}) |\eta_{\cal SE}\rangle = |\psi_{\cal SE}\rangle $, it is envariant under $u_{\cal S}$. For such
composite states one can rigorously establish that the local state of ${\cal S}$ remains unaffected by
$u_{\cal S}$. Thus, for example, the phases of the coefficients in the Schmidt decomposition $|\psi_{\cal SE}\rangle = \sum_{k=1}^N a_k |s_k\rangle |\varepsilon_k\rangle $ are envariant, as the effect of $u_{\cal S}=\sum_{k=1}^N \exp(i \phi_k)|s_k\rangle \langle s_k| $ can be undone by a
{\it countertransformation} $u_{\cal E}=\sum_{k=1}^N \exp(-i \phi_k)|\varepsilon_k\rangle
\langle \varepsilon_k| $ acting solely
on the environment. This envariance of phases implies their
irrelevance for the local states -- in effect, it implies decoherence. Moreover, when the absolute values of the
Schmidt coefficients are equal (as in the figure above), a swap
$\ket \spadesuit \bra \heartsuit + \ket \heartsuit \bra \spadesuit$ in $\cS$ can be undone by a
`counterswap' $\ket \clubsuit \bra \diamondsuit + \ket \diamondsuit \bra \clubsuit$ in $\cE$.
So, as can be established more carefully \cite{Zurek05},  $p_\spadesuit = p_\heartsuit=\frac 1 2$
follows from the objective symmetry of such an entangled state. This proof of equal probabilities is based not on ignorance (as in Laplace's indifference) but on a perfect knowledge of the ``wrong thing'' -- of the global observable that rules out (via quantum indeterminacy) any information about complementary local observables. When supplemented by simple counting, envariance leads to Born's rule \cite{Zurek03a,Zurek05,Zurek11,Zurek03b}.}
\label{qcards}
\end{figure*}

We start our derivation of the Born's rule with the case of equal probabilities: When the Schmidt
coefficients are equal, symmetries of entanglement force one to conclude that the probabilities
must be also equal. The crux of the proof is that, after a swap on the system, the probabilities of
the swapped states must be equal to the probabilities of their new partners in the
Schmidt decomposition (which did not yet get swapped). But -- when the coefficients are equal -- a swap on
the environment restores the original states. So the probabilities must be the same as if the swap
never happened. These two requirements (that a swap exchanges probabilities, and that it does not
change them) can be simultaneously satisfied only when probabilities are equal.

In quantum physics one seeks probability of measurement outcome starting from a known state of $\cS$ and a ready-to-measure state of the apparatus pointer $\cA$. The entangled state of the whole is pure, so (at least prior to the decoherence by the environment) there is no ignorance in the usual sense. In particular, repeatability postulate (iii) assures that the presence of a known state can be verified -- confirmed by a suitable measurement.  Thus, repeatability implies certainty, providing a normative case of probability.

{\it Envariance} in a guise slightly different than before (when it accounted for decoherence) implies mutually exclusive outcomes with certifiably equal probabilities:
Suppose $\cS$ starts as $\ket \rightarrow =\frac { \ket \uparrow + \ket \downarrow} {\sqrt 2}$, so interaction with $\cA$ yields $\frac { \ket \uparrow \ket {A_\uparrow}  + \ket \downarrow \ket {A_\downarrow}} {\sqrt 2}$, an {\it even} (equal coefficient) state 
(and we skip normalization below to save on notation).

The unitary {\it swap} $ \kb \uparrow \downarrow + \kb \downarrow \uparrow$ permutes states in $\cS$:


\[
\tikz[baseline]{
            \node[fill=gray!20,anchor=base] (t1)
            {$|\uparrow \rangle$};
        } 
| {A_\uparrow} \rangle 
+
\tikz[baseline]{
            \node[fill=gray!20,anchor=base] (t2)
            {$|\downarrow \rangle$};
        }
|{A_\downarrow} \rangle
\quad \longrightarrow \quad
| \downarrow\rangle |{A_\uparrow} \rangle + |\uparrow  \rangle | {A_\downarrow} \rangle . \ \eqno(3a)
\]
\begin{tikzpicture}[overlay]
        \path[->] (t1) edge [bend left=40] (t2);
        \path[->] (t2) edge [bend left=40] (t1);
\end{tikzpicture}
%


\noindent 
After the swap $ \ket \downarrow $ is as probable as $\ket {A_\uparrow}$ was (and still is), and $\ket \uparrow$ as $\ket {A_\downarrow}$.  Probabilities in $\cA$ are unchanged 
(as $\cA$ is untouched) so $p_\uparrow$ and $p_\downarrow$ must have been swapped. To prove equiprobability we now 
swap records in $\cA$:


%
\[
|\downarrow\rangle \tikz[baseline]{
            \node[fill=gray!20,anchor=base] (t3)
            {$| {A_\uparrow}\rangle$};
        } 
+
|\uparrow \rangle
\tikz[baseline]{
            \node[fill=gray!20,anchor=base] (t4)
            {$|{A_\downarrow} \rangle$};
        }
\quad \longrightarrow \quad
|\downarrow\rangle |{A_\downarrow} \rangle| + |\uparrow \rangle | {A_\uparrow}\rangle . \ \eqno (3b)
\]
\begin{tikzpicture}[overlay]
        \path[->] (t3) edge [bend left=40] (t4);
        \path[->] (t4) edge [bend left=40] (t3);
\end{tikzpicture}
%


\noindent  Swap in $\cA$ restores the pre-swap state $ \ket \uparrow \ket {A_\uparrow}  + \ket \downarrow \ket {A_\downarrow}$ without touching $\cS$, so (by fact 3) the local state of $\cS$  is also restored (even though, by fact 1, it could not have been affected by the swap of Eq. (3a)).
Hence (by fact 2), all predictions about $\cS$, {\it including probabilities}, must be the same!
The probabilities of $\ket \uparrow$ and $ \ket \downarrow$, (as well as of $\ket {A_\uparrow} $ and $\ket {A_\downarrow}$) are exchanged yet unchanged. Therefore, 
they must be equal. Thus, in our two state case 
$p_\uparrow=p_\downarrow= \frac 1 2$. For $N$ envariantly equivalent alternatives, $p_k= \frac 1 N\ \forall k$.

Getting rid of phases beforehand was crucial:
Swaps in an isolated pure states will, in general, change the phases, and, hence, change the state.  For instance, $\ket \spadesuit + i \ket \heartsuit$, after a swap $\ket \spadesuit \bra \heartsuit + \ket \heartsuit \bra \spadesuit$, becomes
$i \ket \spadesuit + \ket \heartsuit$, i.e., is orthogonal to the pre-swap state. The crux of the proof of
equal probabilities was that the swap does not change anything {\it locally}. This can be established
for entangled states with equal coefficients but -- as we have just seen -- is simply not true for a pure state of just one system.

In the real world the environment will become entangled (in the course of decoherence)
with the preferred states of the system of interest (or with the preferred states of the apparatus pointer).
We have just seen how postulates (i) - (iii) lead to preferred sets of states. We have also pointed
out that -- at least in idealized situations -- these states coincide with the familiar pointer states
that remain stable despite decoherence. So, in effect, we are using the familiar framework of
decoherence to derive Born's rule. Fortunately, as we have seen, it can be analyzed without
employing the usual (Born's rule - dependent) tools of decoherence (reduced density matrix and trace). 

So far we have only explained how one can establish the equality of probabilities for the
outcomes that correspond to Schmidt states associated with coefficients that differ at most by a phase. This is not yet Born's rule. However, it turns out that this is the hard part of the proof: Once such equality is established, a simple counting argument 
leads to the relation between probabilities and unequal coefficients \cite{Zurek03a,Zurek05,Zurek11,Zurek03b}.

Thus, for an uneven state $\ket {\phi_{\cS\cA}}=\alpha\ket \uparrow \ket {A_\uparrow}  + \beta \ket \downarrow \ket {A_\downarrow}$ swaps on $\cS$ and $\cA$ yield $\beta \ket \uparrow \ket {A_\uparrow}  + \alpha \ket \downarrow \ket {A_\downarrow}$, and not the pre-swap state, so $p_\uparrow$ and $p_\downarrow$ are not equal. However, the uneven 
case reduces to equiprobability via {\it finegraining}, so envariance, Eq. (8), yields Born's rule, 
$p_{s|\psi}=|\bk s \psi|^2$, 
in general.

To this end, let $\alpha \propto \sqrt \mu, ~ \beta \propto \sqrt \nu $, where $\mu, \nu$ are natural numbers (so the squares of $\alpha$ and $\beta$ are commensurate). To finegrain, we change the basis; $\ket {A_\uparrow}=\sum_{k=1}^\mu \ket {a_{ k}}/\sqrt \mu$, and $ \ket {A_\downarrow}=\sum_{k=\mu+1}^{\mu+\nu} \ket {a_{ k}}/\sqrt \nu$, in the Hilbert space of $\cA$:
\vspace{-0.1cm}
$$
 \ket {\phi_{\cS\cA} } \propto \sqrt \mu ~ \ket \uparrow \ket {A_\uparrow} + \sqrt \nu ~ \ket \downarrow \ket {A_\downarrow} = $$
\vspace{-0.5cm}
 $$
= \sqrt \mu ~ \ket \uparrow \sum_{k=1}^\mu \ket {a_{ k}}/\sqrt \mu + \sqrt \nu ~ \ket \downarrow\sum_{k=\mu+1}^{\mu+\nu} \ket {a_k}/\sqrt \nu \ . 
$$ 
We simplify, and imagine environment decohering $\cA$ in a new orthonormal basis. That is, $\ket {a_k}$ correlate with $\ket {e_k}$ so that;
$$
\ket {\Phi_{\cS\cA\cE} } \propto\sum_{k=1}^\mu \ket {\uparrow {a_k}} \ket {e_k}+ \sum_{k=\mu+1}^{\mu+\nu}\ket {\downarrow {a_k} } \ket {e_k} 
$$
as if $\ket {a_k}$ were the preferred pointer states. Now swaps of $\ket {\uparrow {a_k}}$ with $\ket {\downarrow {a_k} } $ can be undone by counterswaps of the corresponding $\ket {e_k}$'s.
Counts of the finegrained equiprobable $(p_k=\frac 1 {\mu + \nu})$ alternatives labelled with $\uparrow$ or $\downarrow$ lead to Born's rule:
$$ p_\uparrow = \frac \mu {\mu + \nu} =|\alpha|^2, \ \ \ \ p_\downarrow = \frac  \nu {\mu + \nu} = |\beta|^2 . 
$$
Amplitudes `got squared' as a result of Pythagoras' theorem (the Euclidean nature of Hilbert spaces). Continuity settles the case of incommensurate $|\alpha|^2$ and $ |\beta|^2$\footnote{When the probabilities are incommensurate, one uses sequences of states with commensurate probabilities (deduced from the Schmidt coefficients as above) that converge, from above and below, on these incommensurate probabilities: Such sequences ``bracket'' the incommensurate probabilities. We assume that probabilities are continuous functions of quantum states. Therefore, incommensurate probabilities can be deduced using this ``Dedekind-cut -- like'' strategy we have just outlined.}. 

\subsection{Discussion of envariant derivation of Born's rule}

In physics textbooks Born's rule is a postulate. Using entanglement we derived it here from the quantum core axioms. Our reasoning was purely quantum: Knowing a state of the composite classical system means knowing state of each part. There are no entangled classical states, and no objective symmetry to deduce classical equiprobability, the crux of our derivation. Entanglement -- made possible by the tensor structure of composite Hilbert spaces, introduced by the composition postulate (o) -- was key. The appeal to symmetry -- a strategy that is subjective and suspect in the classical case -- becomes rigorous thanks to objective envariance in the quantum case. 
Born's rule, introduced by textbooks as postulate (v), follows. 

Relative frequency approach (found in many probability texts) starts with ``events''. It has not led to successful derivation of Born's rule (see e.g. assessment by Weinberg \cite{Weinberg}). We used entanglement symmetries to identify equiprobable alternatives. However, employing envariance one can deduce frequencies of events by considering $M$ repetitions (i.e., $(\alpha\ket \uparrow \ket {A_\uparrow}  + \beta \ket \downarrow \ket {A_\downarrow})^{\otimes M}$) of an experiment, and deduce departures that are also expected when $M$ is finite. Moreover, one can even show the inverse of Born's rule: that is, one can demonstrate that the amplitude should be proportional to the square root of frequency \cite{Zurek11}.

As the probabilities are now in place, one can think of quantum statistical physics. One could establish its foundations using probabilities we have just deduced. But, as we shall now see, there is an even simpler and more direct approach that arrives at the microcanonical state without the need to invoke ensembles and probabilities. 
The basic idea is to regard an even state of the system entangled with its environment as the microcanonical state. This is a major conceptual simplification: One can get rid of the artifice of invoking infinite collections of similar systems to represent a state of a single system in a manner that allows one to deduce relevant thermodynamic properties.

Our approach is based on envariance. Before we go on, we briefly return to the experimental status of envariance. There are now four experiments. Two are ``real'' laboratory experiments. Thus, Ref. \cite{Laflamme} use Hong-Ou-Mandel effect \cite{HOM} to verify envariance of entangled and spatially separated photons (e.g., they carry out in the laboratory the ``thought experiment'' depicted in Fig. 1). Ref. \cite{Ebrahim} employs tools of atomic physics to ``experimentally unpack'' and confirm several quantum properties used in the envariant derivation of Born's rule. Quantum computers -- IBM 5 qubit and 16 qubit processors -- are used to manipulate quantum states of individual  Josephson junction qubits by Deffner \cite{Deffner} and by Ferrari and Amoretti \cite{Ferrari} in an imaginative use of quantum information processing to test fundamental quantum predictions -- of the symmetries of entanglement that underlie our derivation of Born's rule and our discussion of the foundations of quantum statistical physics\cite{sample}.

\section{Equilibrium from the symmetries of entanglement}

We consider again a quantum system $\cS$ entangled with its environment $\cE$. The state of the composite $\cS \cE$ is pure, and can be represented by Schmidt decomposition:
$$ | \Psi_{\cal SE}\rangle \propto \sum_{k=1}^K \alpha_k
|s_k\rangle |\varepsilon_k\rangle \ . \eqno(4a) $$
In the above, $\alpha_k$ are Schmidt coefficients. The two bases, $\{\ket {s_k}\}$ in the Hilbert space of the system $\cal H_\cS$, and $\{\ket {\varepsilon_k}\}$ in the Hilbert space $\cal H_\cE$ of the environment, are orthonormal. 

The phases of the Schmidt coefficients can be rotated by acting on $\cS$ alone with the unitary ${\bf U}_\cS^\phi=\sum_k e^{i \phi_k} \ket{s_k} \bra{s_k} $.
This phase rotation can be undone by ``counter-rotation'', a unitary that acts only on the environment, ${\bf U}_\cE^{-\phi}=\sum_k e^{-i \phi_k} \ket{\varepsilon_k} \bra{\varepsilon_k} $. As the state of the whole composite $\cS\cE$ (that might have been affected by ${\bf U}_\cS^\phi$) is restored by ${\bf U}_\cE^{-\phi}$, it follows that the state of $\cS$ is also restored. 

This is a simple example of {\it entanglement-assisted invariance} or {\it envariance}. Envariance shows that the state of $\cS$ alone does not depend on the phases of Schmidt coefficients, as the state of the whole $\cS\cE$ (hence, also the state of $\cS$) is restored without acting on $\cS$. As we have already seen, this phase independence provides, as a corollary, a very fundamental view  \cite{Zurek03a,Zurek03b,Zurek05,Zurek11} of the essence of decoherence \cite{GJK+96a,Zurek03b,Schlosshauer,Zurek14}.

\subsection{Microcanonical equilibrium}

The microcanonical ensemble corresponds to an {\it even} quantum state represented by Schmidt decomposition when the absolute values of all the coefficients are equal \cite{DeffnerZurek2016}:
$$|\bar \Psi_{\cal SE}\rangle \propto \sum_{k=1}^K e^{i \phi_k}
|s_k\rangle |\varepsilon_k\rangle \ . \eqno(4b)$$
We shall now show that the state of the system alone does not -- cannot -- evolve, exhibiting key feature of an equilibrium state in a single system. 

This immunity to change under any unitary evolution is easily seen: Any unitary evolution operator 
$$ {\bf U}_{\cal S}(\{ \tilde s_k\} \rightleftharpoons \{s_k\}) =
\sum_{|s_k\rangle \in {\cal H_S}} |\tilde s_k \rangle \langle s_k| \eqno(5a)$$ 
acting on the system can be undone by a corresponding counter-evolution operator:
$$ {\bf U}_{\cal E}(\{\tilde\varepsilon_k\}\rightleftharpoons\{\varepsilon_k\})
= \sum_{\ket {\varepsilon_k} \in {\cal H_E}} |\tilde\varepsilon_k \rangle
\langle \varepsilon_k| \eqno(5b)$$
acting on the environment. Above, $\{|s_k\rangle\}$ and $\{|\tilde s_k\rangle\}$ are two orthonormal basis sets that span the same ${\cal H_S}$ (or possibly, just the same {\it even subspace} ${\bar {\cal H}_{\cal S}}$ of ${\cal H_S}$). 

\vspace{2mm}
{\bf Theorem}: When the state of whole $\cS\cE$ can be restored without undoing the evolution of $\cS$, it follows that ${\bf U}_{\cal S}$ has not induced any evolution in $\cS$.

\vspace{2mm}

{\bf Proof}: ${\bf U}_{\cal S}(\{ \tilde s_k\} \rightleftharpoons \{s_k\})$ could (and, generally, would) affect the state of $\cS$. However, when $\ket {\bar \Psi_{\cal SE}}$ is even, its effect can be undone by the transformation ${\bf U}_{\cal E}(\{\tilde\varepsilon_k\}\rightleftharpoons\{\ket {\varepsilon_k} \})$ that does not act at all on $\cS$. The even state of the whole, $\ket {\bar \Psi_{\cal SE}}$, is then restored by ${\bf U}_{\cal E}(\{\tilde\varepsilon_k\}\rightleftharpoons\{\ket {\varepsilon_k} \})$. Therefore, by fact 3, the local state of $\cS$ is also restored. However, by fact 1, it could not have been affected by ${\bf U}_{\cal E}(\{\tilde\varepsilon_k\}\rightleftharpoons\{\ket {\varepsilon_k} \})$ that acted only on $\cE$. It therefore follows (by fact 2) that measureable properties of $\cS$  could not have been altered by ${\bf U}_{\cal S}(\{ \tilde s_k\} \rightleftharpoons \{s_k\})$ -- local state of $\cS$ could not have evolved -- when the whole was an even $\ket {\bar \Psi_{\cal SE}}$. QED.
\vspace{2mm}

To establish this indifference to evolutions it suffices to show:
\vspace{2mm}
 
{\bf Lemma:} When $|\bar \Psi_{\cal SE}\rangle$ is even, and the states $\{|s_k\rangle\}$, $\{|\tilde s_l\rangle\}$ 
are orthogonal for each of these two bases (as demanded by unitarity of ${\bf U}_{\cal S}(\{ \tilde s_k\} \rightleftharpoons \{s_k\})$, a unitary counter-evolution operator $ {\bf U}_{\cal E}(\{\tilde\varepsilon_k\}\rightleftharpoons\{\varepsilon_k\})$ exists. 
\vspace{2mm}

{\bf Proof}: Any  unitary can be expressed as
$ {\bf U}_{\cal S}(\{ \tilde s_k\} \rightleftharpoons \{s_k\}) =
\sum_{|s_k\rangle \in \bar {\cal H_S}} |\tilde s_k \rangle \langle s_k|$, a ``partial swap''.
This is the obvious generalization of the simple swap of Eqs. (3a,b). The partial swap
$ {\bf U}_{\cal S}(\{ \tilde s_k\} \rightleftharpoons \{s_k\})$ can be undone
by the corresponding {\it partial counterswap} of the Schmidt partners of
the swapped pairs of states. This is because $\ket {\bar \Psi_{\cal SE}}$ (state envariant under complete swaps) must be even -- i.e., must have a form $|\bar \Psi_{\cal SE}\rangle \propto \sum_{k=1}^K e^{i \phi_k}
|s_k\rangle |\varepsilon_k\rangle$
where $K=Dim(\bar {\cal H_\cS})$. Basis $\{|\tilde s_l\rangle\}$ spans
the same subspace $\bar {\cal H}_{\cal S}$.  Therefore,
$$|\bar \Psi_{\cal SE}\rangle \propto \sum_{l=1}^K |\tilde s_l \rangle
(\sum_{k=1}^K e^{i \phi_k} \langle \tilde s_l |s_k\rangle 
|\varepsilon_k\rangle)
= \sum_{l=1}^K |\tilde s_l\rangle |\tilde \varepsilon_l\rangle\ . $$
Given that $\{|\varepsilon_k\rangle\}$ are Schmidt, one can 
verify that $\{|\tilde \varepsilon_l\rangle = (\sum_{k=1}^K e^{i \phi_k} \langle \tilde s_l |s_k\rangle 
|\varepsilon_k\rangle)\}$ are orthonormal, and, therefore,
the expansion on RHS above is also Schmidt. Consequently, the counter-evolution operator
${\bf U}_{\cal E}(\{\tilde\varepsilon_k\}\rightleftharpoons\{\ket {\varepsilon_k} \})$, Eq. (5b)
-- that is, the desired partial counterswap -- exists. This establishes envariance under arbitrary unitaries /
partial swaps. QED.

\noindent{We conclude that unitaries acting on $\cS$ alone do not affect its state, Eq. (4b), when it is entangled with $\cE$. }
The microcanonical equilibrium can be still represented by the unit density matrix, $ \rho_\cS \propto {\bf 1}_\cS$. One can obtain it by tracing out $\cE$. However, tracing out is averaging, so the physical interpretation of reduced density matrices invokes Born's rule. 

We have arrived at the microcanonical state without invoking Born's rule (not to mention ensembles). Symmetries of entanglement were enough.
As the probabilities of all the states in any orthocomplete basis are the same, the microcanonical equilibrium state does not support any density gradients or flows. Indeed, one can reach the conclusion about the absence of flows more directly: Envariance eliminates local phases in the system, and without phase gradients in a quantum system there are no probability currents -- no flows.

\subsection{Canonical equilibrium}

Once microcanonical state is in place, the canonical equilibrium and the Boltzmann distribution (probability of occupation of energy levels) are obtained by regarding system $\cS$ that is in microcanonical equilibrium (because of its entanglement with the environment $\cE$) as a composite that comprises the system of interest $\mathfrak s$ and its heat bath $\mathfrak b$. This is the usual textbook strategy \cite{Toda}: The microcanonical equilibrium represents a degenerate energy eigenstate, so the levels in $\mathfrak s$ are occupied with probabilities obtained under the constraint that the total energy of $\mathfrak s$ plus its heat bath $\mathfrak b$ adds up to the energy of the microcanonical state of $\cS$. When one assumes that the number of energy levels in $\mathfrak b$ is large compared to the number of energy levels in $\mathfrak s$, and the interaction between the two is negligible compared to their self-Hamiltonians (standard textbook assumptions) Boltzmann distribution over energy levels in $\mathfrak s$ follows \cite{DeffnerZurek2016}. The inverse temperature $\beta$ appears (as usual) as the Lagrange multiplier\footnote{Of course, one can consider distribution over the energy eigenstates of the system without assuming that the environment / heat bath are effectively infinite.}.

There are two parts in our derivation of the canonical state. Part I results in the microcanonical state. It is based on envariance and is very different from textbook arguments -- it does not require ensembles or any other excuses such as ergodicity (unavoidable in the classical context of Newtonian physics) to capture the essence of the microcanonical equilibrium. Rather,  a single system entangled with its environment is enough. 

Thus, quantum mechanics gets rid of the artificial (if succesfull) statistical ``model'' of thermodynamic equilibrium -- an infinite ensemble of systems to ``stand in'' for a single system. Part II -- the derivation of the canonical equilibrium from the microcanonical state -- 
is ({\it mutatis mutandis}) ``textbook''. In the end (retracing to some extent the derivation of Born's rule from envariance) one obtains the density matrix of a {\it single} system, with the Boltzmann energy distribution, but without ensembles, ergodicity, etc. 

\subsection{Discussion}

One can also compare our derivation of canonical equilibrium to the derivation of Born's rule when absolute values of the Schmidt coefficients are not equal. There  \cite{Zurek03a,Zurek03b,Zurek05,Zurek11,Zurek14} one considers states of the form of Eq. (4a). As we have seen in the previous section, when the squares $|\alpha_k|^2$ of the coefficients are commensurate, one can ``finegrain'' the entangled state by appending an extra environment $\cE^\prime$ correlated with the primary environment $\cE$ in such a way that all the triplet states $\ket {s_k} \ket {\varepsilon_{k_j}} \ket {\varepsilon_{k_j}^\prime}$ have the same coefficients, and, therefore, same probabilities. The probability of $\ket {s_k}$ is then proportional to the number of states $K_j \propto |\alpha_k|^2$ in the superposition $\sum_j^{K_j} \ket {\varepsilon_{k_j}} /  \sqrt{K_j}= {\ket {\varepsilon_k}}$ that yields the original Schmidt states. For incommensurate $|\alpha_k|^2$ one recovers Born's rule by appealing to continuity.

A similar situation naturally arises in the microcanonical setting we have just described. The state we have used to represent the microcanonical equilibrium of $\cS$ is already even, so there is no need to finegrain. The canonical state is obtained by coarse-graining (so in a sense we are using the finegraining strategy that led to Born's rule, but doing it ``in reverse''). The goal is to count the states in $\mathfrak {s b}$ that correspond to energy $E_k$ in $\mathfrak s$. That count yields \cite{DeffnerZurek2016} the Boltzmann factor when the total number of states is maximized subject to the constraint that the energy of $\mathfrak {s b}$ is fixed and the number of states in the heat bath $\mathfrak b$ is large compared to the number of states in the system $\mathfrak s$.

It is straightforward to represent the canonical equilibrium state in the system $\mathfrak s$ as a Schmidt decomposition of a pure state in an enlarged Hilbert space:
$$ \ket {\Psi_{\mathfrak s \mathfrak E}} = \sum_k e^{-{\frac {\beta E_k} 2} } \ket {\sigma_k} \ket {\varepsilon_k} $$
In the above, $\mathfrak E$ is the composite of $\mathfrak b$ and $\cal E$, $\ket {\sigma_k}$ are the energy eigenstates of $\mathfrak s$, while $\ket {\varepsilon_k}$ are the corresponding states in $\mathfrak E$. Strictly speaking, one would also need to assume that the whole is in a pure state to justify use of the Schmidt decomposition. However, one can also simply regard the above Schmidt decomposition as a purification of the mixed thermal 
state \cite{Umezawa} (as states of $\mathfrak E$ generally do not enter into physical considerations).

\section {Szilard's engine}

Statistical physics was developed in the XIX century to justify thermodynamics when atomic models relied on Newtonian mechanics. The key ingredient employed throughout (although codified only by Gibbs) was the concept of an ensemble. Ensembles represented ``macrostates'' that were subject to thermodynamics. They correspond to collections of microstates, each subject to Newton's laws of motion.

Above we have seen how symmetries of entanglement simplify the foundations of quantum statistical physics, removing the need for ensembles. Our Universe is quantum (as all experiments to date confirm). Thus, in our quantum Universe, we can practice statistical physics without ensembles. Indeed, statistical physics may be statistical precisely because of the probabilities introduced by quantum physics. The ``take home message'' of the above discussion is that one does not need ensembles to deduce thermodynamics: In our quantum Universe a single system entangled with its heat bath is enough.

\begin{figure*}[tb]
\begin{tabular}{l}
\includegraphics[width=6.0in]{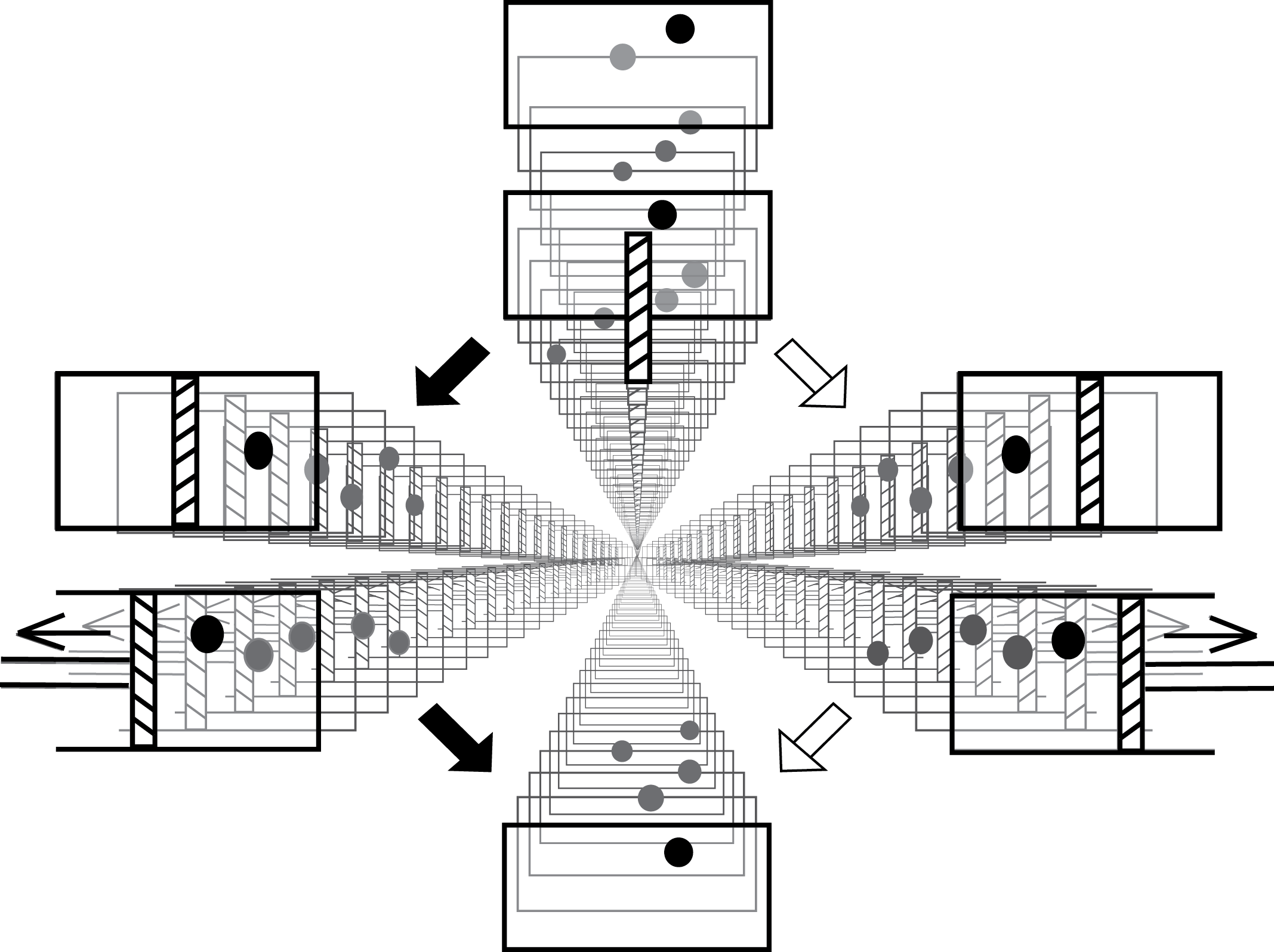}\\
\end{tabular}
\caption{A statistical ensemble representing a single cycle of just one Szilard's engine \cite{Szilard}. To deduce the thermodynamics of the engine, one needs to consider an infinite collection of points in phase space representing the location of the gas particle. Insertion of the partition compresses the gas (in violation of the law of Gay-Lussac) to half of its initial volume. Extraction of work requires information about the side of the container where the particle is trapped. The subsequent expansion of the gas yields work. The second law is safe when the thermodynamic cost of operation of the measuring device is at least as large as the work extracted, as is mandated by Landauer's principle \cite{Landauer,Bennett}.
}
\label{ensemble}
\end{figure*}

While our Universe is quantum, we have persistent illusion that the world we live in is classical. This illusion can be explained with an appeal to decoherence \cite{GJK+96a,Zurek14,Zurek03b,Schlosshauer}. However, ensembles ``work'' -- they have led to useful conclusions and are the staple of textbook discussions of statistical physics. Therefore, despite the awkwardness involved in representing a single system with an infinite ensemble, they have clearly captured an important element of truth. It is interesting to compare, side by side, classical and quantum accounts of thermodynamics of similar systems and to distill the element of truth captured by the concept of ensembles. To pursue this goal we select here the well-known classical Szilard's engine and devise a suitable quantum counterpart. Our discussion brings together several themes (including information and randomness) that are essential for the foundations of both thermodynamics and quantum physics.

Szilard \cite{Szilard} described, in 1929, a thought experiment involving Maxwell's demon operating a single classical molecule heat engine (Fig. 3). It is now familiar to many, so my presentation of Szilard's engine will be brief: The engine consists of a cylinder (volume $V$) that contains a single molecule of gas in contact with a thermal reservoir (temperature $T$). The cycle starts with insertion of a partition in the middle of the cylinder. This divides the volume $V$ into two halves. The classical gas molecule must be either right or left of the partition, so the gas becomes confined to the volume of $V/2$. The partition can be now used as a piston. The work of:
$$
\Delta W = \int_{V/2}^{V} p(v) dv = k_BT \int_{V/2}^{V} dv/v = k_BT\ln2 \eqno(6)
$$
is extracted in course of the isothermal expansion. If this process could be repeated, each cycle would lead to extraction of $k_BT\ln2$ of useful work. 
This {\it perpetuum mobile} would obviously threaten the second law of thermodynamics. 

This threat was the main preoccupation of Szilard. He concluded that the repeated measurements (needed to decide which way the partition should expand when it plays the role of the piston) are possible only if, after each cycle, the state of demon's memory is restored to the ``ready to measure'' state at the free energy cost of $k_BT\ln2$. This anticipated Shannon's definition of information \cite{Shannon}, its connection with thermodynamic entropy and set the stage for Landauer's principle \cite{Landauer,Bennett}.

The focus of our discussion is different: At the instant when the partition is closed, the first law is threatened, as the gas is -- with no work expenditure from the agent operating the engine -- compressed to half its initial volume. This process should take no less than $\Delta W = k_BT\ln2$ of work. 
Thus, even if one agrees with Szilard's (plausible) conclusion that -- over the whole cycle -- useful work cannot be repeatably extracted, the insertion of the partition appears to increase e.g. Helmholtz free energy. So, one might argue that the threat posed by Szilard's classical engine is also to the first, rather than just to the second, law of thermodynamics. Of course, the agent must measure to find out where is the molecule to liberate this extra free energy. Still, in this classical setting, the decrease of entropy caused by the measurement is subjective -- the state of the engine (the location of the molecule) does not change. It is just that the agent finds out where it is. 

One is now confronted with an issue reminiscent of these encountered in the context of quantum measurements: Is the state of the Universe one should focus on the ``true'' classical state (with the molecule on just one side of the partition) or is it the state observer must use (with both ``branches'' in Fig. 3) to infer the laws of thermodynamics? We shall see that quantum engine we are about to consider sheds a new light on this conflict between what is and what is known, and that -- in this context -- our definition of equilibrium is helpful.

The origin of the problem is clear: Two distinct idealizations of gas are used. Thermodynamic idealization -- Gay-Lussac law, $pV= k_BT$ -- is used above to compute the extracted work. However, as the partition divides the cylinder into two halves, one appeals to the ``single classical molecule'' -- a very different model of gas. The ensemble account of engine cycle helps defuse (but does not completely resolve) the conflict. The problem arises as the partition is closed: One can certainly believe that the appropriate averaging over the molecule locations and momenta within the ensemble will result in the law of Gay-Lussac, but effortless compression to half the volume by the closing of the partition violates this very law. 

Jauch and Baron \cite{JauchBaron}, in their critique of Szilard's engine, noted this inconsistency. I shall not dwell on it here: I believe their criticism -- illegality of the use of two distinct idealizations of gas in the same argument -- is largely valid. The ultimate problem is, however (as noted some time ago \cite{ZurekSzilard}), classicality (hence, locality) of the molecule that necessarily ``gets stuck'' on a single side of the partition. Slow insertion of partition should follow a sequence of equilibrium states and should be thermodynamically reversible. Yet, the final closing of the partition confines the classical molecule into half its initial volume, and this far-from-equilibrium consequence happens without any expenditure of work.
Moreover, a sufficiently slow process should yield a sequence of (near)equilibrium states. Instantaneous compression of the gas does not.\footnote{There is a subtlety here: Slow has to be defined with respect to some other relevant timescale. One could argue that the relevant timescale in this example is the time it takes the molecule to ``switch sides'', and that this timescale becomes infinitely long as the opening left by the partly inserted partition becomes sufficiently small. We shall not delve into this issue in any more detail. We only note that it does not resolve the main problem we are faced with -- the conflict between the particle being stuck on just one side of the partition and the consequent effortless compression, in violation of the law of Gay-Lussac.}.



\section{Quantum one-molecule engine} 

In spite of the concern about the effortless compression of the gas
I believe Szilard's conclusion is valid -- information the agent does not have prevents him from extracting useful work. 
This observation has been by now encapsulated as Landauer's principle \cite{Landauer} (see especially the discussion by Bennett \cite{Bennett}). However, I am also convinced that this conclusion anticipates the role information plays in quantum theory, the role that ultimately translates into the approach to foundations of statistical physics and thermodynamics discussed here and in \cite{DeffnerZurek2016}: We shall see -- analyzing the quantum engine introduced in \cite{ZurekSzilard} -- that the role of information transfer is central in quantum engines, in that the measurement (and not the insertion of the partition) is responsible for the compression of the gas. Thus, the far-from-equilibrium process is there, but it is precipitated by measurement. This direct link between compression and measurement is tied to the absence of ensembles: In contrast to the classical setting, the agent cannot just recognize what already exists and has a well-defined state mandated by Newtonian dynamics. Rather, measurement -- information flow -- is crucial in creating the world the agent will inhabit.

\subsection{Quantum molecule in a box}
The cycle of the quantum engine starts when the analogue of the partition -- potential barrier erected in the middle of $L$-sized infinitely deep well containing a single quantum particle of mass $m$ -- is introduced in the ``cylinder'' of our quantum engine (Fig. 4). The barrier is thin, $d \ll L$. Its height $U$ increases sufficiently slowly to be either thermodynamically or adiabatically reversible. One can make it arbitrarily thin and infinitely high, and in that limit it is given by the Dirac $\delta$ function, as was recently discussed in a related context \cite{Koreans}.

\begin{figure*}[tb]
\begin{tabular}{l}
\includegraphics[width=5.0in]{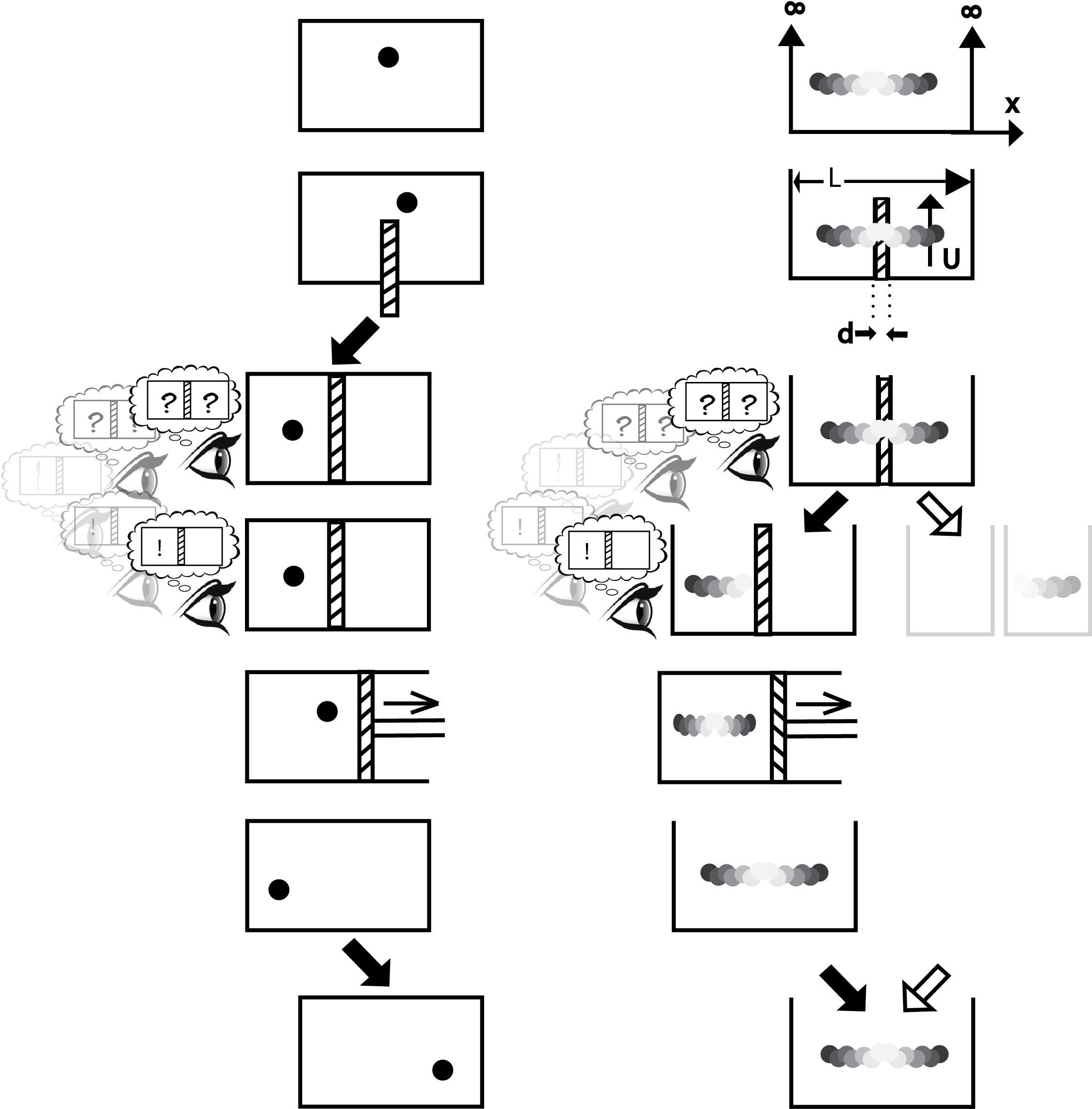}\\
\end{tabular}
\caption{A single cycle of Szilard's engine and of the quantum engine discussed here, compared and contrasted. The main difference is in the role of information: In the classical case (on the left), measurement simply reveals what has already happened. (The ensemble seen in Fig. 3 does not correspond to a classical microstate of a single system, but, rather, to the agent's ``model'' of its macrostate that reflects his ignorance.) In the quantum case (on the right) the molecule is simultaneously on both sides of the partition, until its location is determined by the observer. (The significance of the unobserved branch is a central subject of the discussions on the interpretation of quantum theory, but not of this paper.) The quantum-classical distinction is closely related to the origin of quantum mixtures. They arise as a result of entanglement or simply correlations with the bath (and, more generally, with the rest of the Universe). 
}
\label{engines}
\end{figure*}

The initial square well one-dimensional container has the well-known set of eigenvalues and eigenfunctions:
$$
E_n =  n^2 \pi^2 \hbar^2/(2mL^2)  = \epsilon n^2  \eqno(7)
$$
\vspace{-0.2in}
\renewcommand\theequation{8}
\begin{equation}
\langle x | \psi_n \rangle = \left\{ 
\begin{array}{ll}
(2/L)^{1/2} \cos 2 \pi n x/L  & \mbox{for $n = 2k+1$} \\
(2/L)^{1/2} \sin 2 \pi n x/L  & \mbox{for $n = 2k$} 
\end{array} 
\right.
\end{equation}

The particle inside the potential well is our system $\mathfrak s$. It interacts weakly with  thermal bath $\mathfrak b$. Therefore, the states of $\mathfrak s$ and $\mathfrak b$ are correlated (and we can think of $\mathfrak s$ as entangled with the enlarged environment $\mathfrak E$ consisting of $\mathfrak b$ and the original $\cE$). We assume (in accord with the above discussion) that $\mathfrak s$ is in canonical equilibrium at temperature $k_B T= 1/\beta$, so its state is given by the density matrix:
$$
\rho_{\mathfrak s} = Z^{-1} \sum_n \exp(- \beta \epsilon n^2) |\psi_n \rangle \langle \psi_n | \eqno(9)
$$
The density matrix $\rho_{\mathfrak s} $ is a correct quantum description of a {\it single} quantum system: There is no ensemble -- ${\mathfrak s}$ is in equilibrium with ${\mathfrak b}$, and they may be jointly entangled with the rest of the Universe. So, as our discussion of the origin of Born's law established, this single quantum gas particle is in a mixed state.

The thermodynamic properties of ${\mathfrak s}$ are characterized by its partition function: 
$$
Z = \sum_{n=1}^\infty \exp(- \beta \epsilon n^2) = \sum_{n=1}^\infty \varsigma^{n^2} \eqno(10)
$$
We are interested in the case of high temperatures (where we can compare our quantum engine with that of Szilard). This simplifies the analysis as for $1/2 < \varsigma < 1$ the partition function is approximated by
$Z = \frac{1}{2} \left(\sqrt{\pi/|\ln \varsigma|} -1 \right)$. We can use a still simpler approximation:
$$
Z = (\pi / \epsilon \beta)^{1/2} / 2 = L / \left(h^2/2 mk_BT\right)^{1/2} \eqno(11)
$$
valid for $\epsilon \ll k_B T$ This is  the partition function of the one-dimensional Boltzmann gas. Its three - dimensional version,
$Z = L_x L_y L_z/(h^2/2 \pi mk_BT)^{3/2}$
may be even more familiar.

It is straightforward to verify that internal energy, pressure, entropy, and all other thermodynamic quantities match these of the one-molecule gas used in the classical Szilard's engine. However, now its thermodynamic properties are defined without an ensemble: The density matrix, partition function, etc., describe a {\it single} quantum particle ${\mathfrak s}$. If it was isolated from the rest of the world, it could persist forever in an arbitrary superposition of its energy eigenstates. As it interacts with the heat bath ${\mathfrak b}$, its state becomes mixed, $\rho_{\mathfrak s} $. 
So far, this equilibrium via entanglement does not have substantial consequences for our discussion. However, we shall now show that it helps with the violation of the law of Gay-Lussac and avoids appearance of the violation of the first law of thermodynamics pointed out earlier. 

\subsection{Inserting the barrier}

Let us now consider a barrier of width $d\ll L$ that eventually reaches height $U \gg k_BT$ slowly inserted (Fig. 4) in the middle of the box.
The engine can be (i) coupled to, or (ii) decoupled from heat bath while the barrier is introduced, and we assume this is done slowly, to maintain (i) thermodynamic or (ii) adiabatic reversibility.

The presence of the barrier alters the energy levels: those associated with even values of $n$ (and sine wavefunctions) are lifted up only slightly, so the new eigenvalues are: 
$$
E^\prime_{2k} = \epsilon^\prime (2k)^2 + \Delta_k = E_k + \Delta_k \eqno(12a)
$$
where
$\epsilon^\prime = \epsilon L^2 / (L - d)^2$
and
$$
\Delta_k \cong (4 \epsilon^\prime/ \pi) \exp \left(-d\sqrt{2m(U - E_k)/
\hbar} \right) \ \ . \eqno(13)
$$
Eigenvalues corresponding to odd $n$ (and cosine eigenfunctions that have bigger overlap with the barrier) are shifted upwards by
$\Delta E_n \sim (2n+1)\epsilon^\prime$ so that
$$
E^\prime_{2k-1} = \epsilon^\prime (2k)^2 -\Delta_k = E_k -\Delta_k \ . \eqno(12b)
$$
The pair of the eigenvalues $E^\prime_{2k}$, $E^\prime_{2k-1}$ is separated only by $2 \Delta_k \ll E_k$. It can be thought of as a doubly degenerate level $k$, with the degeneracy lifted for finite values of $U$. In the limit $U \rightarrow \infty$ we have two independent wells. For finite $U$, for these levels where $\Delta_k \ll E_k$, eigenfunctions of the whole system can be
reconstructed from the k$^{th}$ eigenfunctions of the left 
$(|L_k\rangle)$ and right $(|R_k\rangle)$ wells:
$$
E_k +\Delta_k \leftrightarrow |\psi_k^+\rangle = (|L_k\rangle - |R_k 
\rangle)  / \sqrt{2} \ , \eqno(14a)
$$
\vspace{-7mm}
$$
E_k -\Delta_k \leftrightarrow |\psi_k^-\rangle = (|L_k\rangle + |R_k 
\rangle)  / \sqrt{2} \ . \eqno(14b)
$$
Alternatively, eigenfunctions of the left and right wells can be expressed
in terms of energy eigenfunctions of the complete Hamiltonian;
\renewcommand\theequation{15{\rm a}}
\begin{equation}
|L_k \rangle  =   (|\psi_k^+\rangle+ | \psi_k^- \rangle) /  \sqrt{2} \ ,
\end{equation}
\vspace{-7mm}
\renewcommand\theequation{15{\rm b}}
\begin{equation}
|R_k \rangle  =   (|\psi_k^-
\rangle- | \psi_k^+ \rangle) /  \sqrt{2} \ .
\end{equation}
For large $U$ these wavefunctions are localized on the left or on the right of the piston.

The density matrix of the whole system can be expressed in either of these bases: 
\renewcommand\theequation{16\rm a}
\begin{equation}
 \tilde{\rho_{\mathfrak s}} = \tilde{Z}^{-1} \sum^\infty_{k=1} \exp(-\beta E_k) 
\left\{\exp(-\beta\Delta_k) |\psi_k^+ \rangle
\langle \psi_k^+|  
+ \exp(\beta\Delta_k) |\psi_k^- \rangle
\langle \psi_k^-| \right\} \ , 
\end{equation}
\vspace{-7mm}
\renewcommand\theequation{16\rm b}
\begin{equation}
 \tilde{\rho_{\mathfrak s}} = \tilde{Z}^{-1} \sum^\infty_{k=1}\exp(-\beta E_k)\left\{\cosh(\beta\Delta_k) 
(|L_k \rangle \langle L_k| + |R_k \rangle \langle R_k|) + \sinh(\beta\Delta_k)
(|L_k \rangle \langle R_k| + |R_k \rangle \langle L_k|)  \right\} \ .
\end{equation}
This is the same density matrix of the same system. In the limit $U \rightarrow \infty$ the splitting disappears, $\Delta_k \rightarrow 0$, so the subspaces corresponding to the distinct energy levels are completely degenerate, and the two equations above simplify:  
\renewcommand\theequation{17\rm a}
\begin{equation}
 \tilde{\rho_{\mathfrak s}} = \tilde{Z}^{-1} \sum^\infty_{k=1} \exp(-\beta E_k) 
\left\{ |\psi_k^+ \rangle
\langle \psi_k^+|  
+ |\psi_k^- \rangle
\langle \psi_k^-| \right\} \ , 
\end{equation}
\vspace{-4mm}
\renewcommand\theequation{17\rm b}
\begin{equation}
 \tilde{\rho_{\mathfrak s}} = \tilde{Z}^{-1} \sum^\infty_{k=1}\exp(-\beta E_k)\left\{ 
(|L_k \rangle \langle L_k| + |R_k \rangle \langle R_k|)  \right\} \ .
\end{equation}
The state of the single quantum system does not exhibit the asymmetry that appeared when the barrier separated the classical container in Szilard's engine, Fig. 3. There the particle was trapped either in the left or on the right half of the container. A single quantum particle can be on both sides of the container. This symmetry in a single system is retained, in quantum physics, without the ensemble (which restored appearance of symmetry ``in the mind of the observer'' even if it was lost in each run of the single engine he operated). There is no threat to the first law of thermodynamics -- the quantum gas is not compressed by the barrier. It will be however compressed by measurement -- a step that is necessary if we are to emulate a cycle of Szilard's engine.

\subsection{Measurement, as reported by the demon} 

We now consider, from the point of view of the demon (the agent who operates the engine), what happens in the rest of the cycle. One important point has been already made: In contrast to the classical Szilard's engine, insertion of the piston alone does not force the molecule to just one side, and, hence, does not create a pressure inequality between the two halves of the container. However, the demon can measure
\renewcommand\theequation{18}
\begin{equation}
\hat{\Pi} = \lambda ({\bf L} - {\bf R})
\end{equation}
to find out if the molecule is on the left or on the right of the partition. 
Here $\lambda \neq 0$ is an arbitrary eigenvalue while;
\renewcommand\theequation{19{\rm a}}
\begin{equation}
{\bf L}= \sum_{k=1}^N |L_k \rangle \langle L_k| \ ,
\end{equation}
\vspace{-3mm}
\renewcommand\theequation{19{\rm b}}
\begin{equation}
{\bf R}= \sum_{k=1}^N |R_k \rangle \langle R_k| \ ,
\end{equation}
and $N$ is sufficiently large, $N^2 \epsilon \beta \gg 1$.

Following the measurement, the density matrix 
becomes either $\rho_L$ or $\rho_R$ where;
\renewcommand\theequation{20{\rm a}}
\begin{equation}
\rho_L = Z_L^{-1}\sum_{k=1}^\infty \exp(-\beta E_k) \cosh 
\beta\Delta_k |L_k \rangle \langle L_k| \ \ ,
\end{equation}
\vspace{-4mm}
\renewcommand\theequation{20{\rm b}}
\begin{equation}
\rho_R = Z_R^{-1}\sum_{k=1}^\infty \exp(-\beta E_k) \cosh 
\beta\Delta_k |R_k \rangle \langle R_k| \ \ .
\end{equation}
Each of these options is selected with the same probability (as may be concluded using envariance of the entangled/correlated state). 

Thus, {\it it is only at the instant of the measurement that the gas is compressed to half its original volume}. One might be tempted to say that the ``collapse of the wavepacket'' is responsible for the compression. However, this is not the collapse of the wavepacket from a superposition: the density matrix of the system $\mathfrak s$ is decohered already before the demon finds out where the particle is. Even when the barrier is finite but large (so the off-diagonal terms have not yet disappeared), $\sum_k \sinh(\beta\Delta_k) (|L_k \rangle \langle R_k| + |R_k \rangle \langle L_k|) $ in Eq. (16b) is small. 

Von Neumann, in his discussion of the measurement process \cite{vonN}, noted separately (i) the conversion of the pure state into a mixture (the part of the measurement process that leads to the entropy increase, is irreversible and, hence, non-unitary) and (ii) the final ``collapse''  leads to the perception of a single outcome. In our case, the pre-collapse density matrix is, in its form, already post-decoherence. The ``collapse'' leads to the information gain by the observer -- to the entropy decrease he will perceive -- and to the compression of the gas. 

This is where our discussion makes contact with questions of interpretation of quantum theory.  Thus, we note that the classical molecule was already on a pre-selected side of the barrier in any given run of the original Szilard's design. The measurement did not affect its state -- it only established where it is, which ``branch'' of the ensemble of Fig. 3 describes the state of his engine in that particular cycle. 

That gain of information was a one-sided affair -- it had no effect whatsoever on the individual classical system. The ensemble was there to represent demon's pre-measurement ignorance -- it was a subjective ``figment of observer's imagination''. Measurement ``collapsed'' the demon's ignorance but left the state of measured system untouched.

By contrast, measurement of the quantum molecule simultaneously alters the state of the demon's records and of the molecule. Some find this ``symmetry'' between the measured and the measurer disturbing. Interpretations have been invented either to find reasons for the asymmetry (e.g., Copenhagen Interpretation \cite{Bohr}) or to extrapolate (sometimes, to an uncomfortable extent) the consequences of the symmetry (Many Worlds Interpretation) \cite{Everett}. 

Our demon-operated quantum engine sheds additional light on the interplay between information and existence in the quantum world: It is clearly impossible to gain information without altering the state of the measured system. However, the acquisition of information is a prerequisite for the extraction of work. The compression of the gas that allows for the extraction of work occurs as a result of measurement. That measurement decreases entropy of the system by one bit. At the instant of measurement there is no corresponding increase of entropy that the agent can perceive. There may be an increase in algorithmic randomness \cite{Zurek89}, but -- at least according to the demon -- the thermodynamic entropy of the world decreases: The state of the gas is better known, and this knowledge is reflected in the state of demon's records. 

We now return to the concern about the violation of the first law of thermodynamics in the classical Szilard's engine. 
The potential to perform useful work is quantified by the free energy $A(T,L)$ that can be calculated from the 
partition function
$
A(T,L) = -k_B T \ln Z(T,L).
$
For our one-molecule engine in 1D this free energy is 
\renewcommand\theequation{21}
\begin{equation}
A = -k_B T \ln [L/(h^2/2 \pi mk_B T)^{1/2}]
\end{equation}
In the quantum case insertion of the barrier has essentially no effect:
\renewcommand\theequation{22}
\begin{equation}
\tilde{A} = -k_B T \ln [(L-d)/(h^2/2 \pi mk_B T)^{1/2}] \ ,
\end{equation}
save for the fact that the volume that contains the molecule decreases by $\frac d L$, so that $\tilde{A} - A  = k_B T \ln (L/(L-d)) \sim O ({\frac d L})$). We take this effect to be negligible: It would disappear
if the cylinder was simultaneously enlarged from $L$ to $L+d$. 

By contrast, measurement confines the particle to half the pre-measurement volume:
\renewcommand\theequation{23}
\begin{eqnarray}
A_L = A_R & = & -k_B T \ln [((L-d)/2) h^2 /2\pi mk_B T)^{1/2}] \nonumber \\
& = & -k_BT(\tilde{A} - \ln 2) \ .
\end{eqnarray}
This results in an increase in the free energy:
\renewcommand\theequation{24}
\begin{equation}
\delta A = A_L - \tilde{A} = A_R - \tilde{A} = k_B T \ln 2 
\end{equation}
that is, in the quantum case, caused not by the introduction of the barrier, but by the measurement (Fig. 4).

In the classical case, confinement of the molecule to half the original volume would have caused similar increase of free energy. Without introducing the ensemble this increase would be ``for free'' -- it arises as a consequence of violation of the law of Gay-Lussac. 
The ensemble helps defuse this appearance of the threat to the first law: The increase of free energy is compensated by the increase of subjective ignorance of the observer. Thus, in analyzing the thermodynamics of the classical Szilard engine, one resorts to the artifice of a statistical ensemble even though the demon is dealing with a single system. 

In discussions of the interpretation of quantum theory much has been made \cite{d'Espagnat} of the distinction between a ``proper mixture'' (collection of identical systems in different states, epitomized by statistical ensembles) and an improper mixture (a single system described by a density matrix as a consequence of its entanglement with its environment for example, as a result of decoherence). One often encounters opinions that only proper mixtures allow for a straightforward statistical interpretation of probability while improper mixtures do not. 

Here we have encountered what is in a sense a reverse of that situation: a microcanonical state based on envariance -- an epitome of such impropriety of an ensemble -- led to a straightforward and consistent description of thermodynamics of our quantum engine while the classical proper mixture representing Szilard's engine resulted in a tension between the fundamental dynamics and its thermodynamic consequences.


\section{Entanglement, Decoherence, Relaxation, and Equilibration}

The aim of this section is to put various processes that are caused or at least influenced by entanglement in the broader context of the approach to equilibrium. This ambitious goal exceeds the scope of our paper, so we shall not be exhaustive -- this is not a review of the second law of thermodynamics. Nevertheless, given the unorthodox definition of equilibrium we have proposed and discussed, a brief overview of how an equilibrium state is attained seems appropriate.

\subsection{Pure Decoherence}

Pure decoherence occurs when the system $\cS$ interacts with its environment through a Hamiltonian that singles out a preferred pointer observable $\hat \Pi $ which also happens to be a constant of motion under the evolution generated by the self-Hamiltonian of the system. In that (rare) case, a decohered state is completely time-independent. However, its density matrix can be very far from that of the equilibrium mixed state. 

Indeed, in models with pure decoherence there can be no approach to equilibrium -- the system will forever ``remember'' its initial state, as the probabilities of the pointer states remain unchanged by pure decoherence.  The Newtonian model of our everyday world is based on a tacit assumption of  ``nearly pure decoherence'', which rests on the fact that decoherence of delocalized, flagrantly non-classical states is very short compared to the timescales responsible for the approach to equilibrium. As a consequence, flagrantly quantum superpositions are quickly wiped out -- they decohere extremely fast. Yet, the subsequent evolution is predictable  \cite{ZP2}, and classical Newtonian dynamics can be used to compute, for example, the orbits of planets in the solar system with great accuracy.

\subsection{Relaxation and Equilibration}

We now distinguish two timescales associated with somewhat different equilibria -- equilibration of the degrees of freedom within the system, and equilibrium with the environment (which in this last case we can regard as a heat bath).

The Hamiltonian of the system (which now, in contrast to pure decoherence, no longer commutes with the pointer observable favored by the interaction with the environment) is responsible for what we shall call relaxation timescale. It is the quantum analogue of the timescale identified by Boltzmann in his classical discussion of the H-theorem, and established rigorously by Kolmogorov and Sinai for classically chaotic systems. It is given by the inverse of the sum of positive Lyapunov exponents also in the quantum chaotic setting \cite{ZP}, even though in the quantum case the ultimate cause of entropy increase is not the arbitrary (but fixed) coarse graining, but the information flow from the system to the correlations between the system and the environment. 

Thus, in a sense, in an open quantum system coarse graining is not an idealization introduced to represent limitations of an observer, but a fact of life -- it stems from the impossibility to enforce perfect isolation: Decoherence introduces effective diffusion in phase space that suppresses details of the evolving state in quantum versions of classically chaotic systems. As a consequence, its von Neumann entropy increases at the rate given by the sum of positive Lyapunov exponents (even though the ultimate cause of entropy increase is the progressive entanglement with the environment) \cite{ZP}.

In finite quantum systems the von Neumann entropy will eventually asymptote to the value set by the volume of the available phase space / Hilbert space. Thus, an open quantum system starting in an arbitrary initial state will first decohere on a (generally fairly short) decoherence timescale. This will result in a rapid increase of entropy to the value set by the probabilities of the pointer states: The information outflow associated with this process is much faster than the rate of exchange of energy. For instance, when decoherence is caused by quantum Brownian motion -- weak coupling to the environment of harmonic oscillators at a finite temperature \cite{HuPazZhang,Schlosshauer,Zurek91} -- the decoherence timescale $\tau_{Dec}$ is faster than the dissipation timescale $1/\gamma$ on which the systems exchange energy with its environment / heat bath by a factor $(\frac {\Delta x} {\lambda_{dB}})^2 $, where $\lambda_{dB}$ is the thermal de Broglie wavelength \cite{Zurek91}.

Thus, entropy production can start with an initial decoherence-induced burst caused by the mismatch between its initial state and the pointer observables favored by decoherence \cite{Yang}. Subsequently, and over a time interval determined by the dynamics, the system will relax into a state described by a density matrix that is approximately independent of time and represents a mixture that fills in the available phase space / Hilbert space. 

This relaxation process is induced by the self-Hamiltonian of the system that does not commute with the interaction Hamiltonian, and, therefore, gradually rotates states on the diagonal of the already decohered density matrix (that commutes with the interaction Hamiltonian) into their superpositions, thus resulting in further decoherence and increase of entropy. In the approximately microcanonical case of nearly perfectly preserved initially well-defined energy the entropy production will stop only when the system is completely entangled with the environment, so the rotation of the states on the diagonal of its density matrix is of no consequence. This state -- in the appropriate limit -- leads to the microcanonical equilibrium, where the initial energy is retained, but the density matrix is (to an excellent approximation) proportional to the identity operator.

This evolution towards the probability density distribution that fills in the available phase space volume happens over a time determined by the self-Hamiltonian of the system. In integrable systems the relaxation will be slow (with logarithmic entropy increase), while in the quantum versions of classically chaotic systems it will happen faster -- entropy will increase approximately linearly with time (with the rate given by the sum of the positive Lyapunov exponents) until saturation \cite{ZP,ZP2}.

The final state results in the local equilibrium, and the term ``equilibration'' is sometimes used to describe such a relaxation process that leads to local equilibrium in classical or quantum systems. As a consequence of relaxation, various degrees of freedom come to maximize the entropy as its mixed state fills up the available phase space subject to conservation laws -- primarily approximate conservation of energy. 

At this stage of evolution the system no longer ``remembers'' its initial state as well as it did after pure decoherence (the detailed probability distribution preserved by pure decoherence is erased by relaxation) but its state may still contain information about the initial value of the quantities that are approximately conserved in spite of the interaction with environment. That is, energy may be conserved on timescales that can be long compared to the timescales that characterize dynamics of the system even though such dynamical timescales are short compared to the timescale of dissipation  $1/\gamma$.

\subsection{Equilibrium with the Heat Bath}

Following the relaxation process the system is already, in a sense, in {\it local} equilibrium -- its internal degrees of freedom are described by a density matrix that maximizes its entropy subject to relevant constraints. However, in general, it would not have yet reached equilibrium with its heat bath / environment: In particular, if the environment is at some fixed temperature, and the sum of the positive Lyapunov exponents is large compared to $\gamma$, the degrees of freedom of the system may be -- after a few Lyapunov timescales -- already approximately in the local equilibrium state, but the system is not yet at the same temperature as its heat bath.

\hocom{There is a significant difference between the quantum and classical equilibration: In the quantum setting we consider the system that is open and is therefore described by the density matrix that maximizes its entropy given the constraints imposed by the quantities that are conserved or at least approximately conserved in spite of decoherence. By contrast, classical chaotic system continues to be represented, at least in principle, by a point in its phase space, although observer, at least under the usual textbook assumptions of coarse-graining and/or ergodicity can effectively represent it by and ensemble. }

While decoherence is caused by the information flow from the system, and relaxation is caused by the redistribution of energy between the degrees of freedom within the system along with the decoherence due to the environment (and the ensuing increase of von Neumann entropy), equilibration with the heat bath describes the energy exchange between the system and the environment. Thus, during decoherence or what we have termed relaxation, the system will primarily exchange information rather than energy with its surroundings: The environment is monitoring the state of the system. By contrast, energy exchange (heat flow) between the system and its environment (heat bath) is the dominant process responsible for the approach to the ultimate thermal equilibrium with the heat bath.

Not much that needs to be said about this last act of equilibration, except that the distinction between relaxation and equilibration we have noted above is a consequence of the division of the whole into the system and the environment / heat bath. When that split is eliminated (as it was done in the derivation of the canonical equilibrium earlier in this paper), we have just one large system. Indeed, our derivation of the canonical equilibrium discussed briefly earlier in this paper (and more carefully in Ref. \cite{DeffnerZurek2016}) relied on this ability to consider, in parallel, the whole and the parts.

\section{Discussion}

Thermodynamics anticipated quantum theory: Planck introduced his constant to explain the black-body spectrum. The main message of our discussion is that, while one can account for thermodynamic behavior using ensembles, they are indispensable only in the setting of the classical, Newtonian world. Their role is to ``cover up'' incessant motion -- to justify equilibrium -- while providing a model for the probabilities, and, hence, for entropy. 

By contrast, in quantum physics probabilities and entropy are there to begin with. A state of a single system entangled or correlated with the environment or a heat bath is represented by a density matrix. Thus, a state of a {\it single} open system suffices to deduce Born's rule for probabilities. 

In a closed system, entropy is conserved under the unitary evolution \cite{ETH}. However, information gained by the environment -- e.g., in course of decoherence -- accounts for the information lost by the observer. We note that this view of the dynamical second law leads (for reasons that are fundamentally quantum) to conclusions about the rate of entropy production \cite{ZP} that were anticipated by the classical discussion of Boltzmann \cite{Boltzmann}. Entropy increases because the system decoheres -- entangles and/or correlates with its environment. Moreover, when the classical motion is chaotic, the resulting entropy production rate is given by the sum of positive Lyapunov exponents determined by its classical Hamiltonian. This is in essence (a Kolmogorov-Sinai version of) Boltzmann's H-theorem. 

The role of the environment in the justification of equilibrium has been also recognized by the discussions that appeal to ``typicality'' \cite{Popescu, Goldstein}. There one considers a collection -- that is, an ensemble -- of states of a composite object consisting of a system and its environment. Since such states are typically entangled, the density matrix of the system alone is typically close to the microcanonical equilibrium. Thus, ensembles and randomness --the ingredients we wanted to do without -- are a part of that approach. Moreover, Born's rule for probabilities is also involved, as tracing (averaging over the environment) is needed to obtain the density matrix of the system from the entangled composite state. 

This differs from both our goals and the tools we relied on. Our starting point is a single entangled composite state and not an ensemble of such states. Our entangled state is ``even'' -- it is chosen to represent equilibrium in a system. Our criterion for equilibrium is the absence of evolution. We establish absence of evolution relying on the symmetries of entanglement alone -- without appeal to Born's rule (which is, after all, on the list of quantum textbook axioms that depend on the controversial ``collapse'', and, hence -- while widely supported by experiments -- calls for a more fundamental justification \cite{Everett}). 

In our wholly quantum engine this quantum view of equilibrium thermodynamics leads to the resolution of the difficulties posed by Szilard's engine. In the classical context closing of the partition is responsible for the compression of the gas, in violation of the law of Gay-Lussac. Thus, a dramatic non-equilibrium effect is induced by an arbitrarily slow process that results in halving of the volume occupied by gas.

Taken at face value, such compression undermines the very foundation of thermodynamics built on the seemingly safe ground of time-independent equilibrium states that can be transformed into one another through sufficiently slow processes. Quantum theory resolves that contradiction: introduction of the potential barrier alone does not confine the gas into either of the two halves. The density matrix describing it need not be interpreted in terms of an ensemble whose role is to represent ignorance of the observer and to provide a ``cover story'' for the probabilities needed to compute entropy.

In the quantum case, the transformation of the state of the gas and the lifting of observer's ignorance happen simultaneously. The density matrix before the measurement has already a post-decoherence form. It is therefore clear that the absence of the off-diagonal terms alone does not induce the confinement of the gas into one of the two halves of the container. Therefore, decoherence does not precipitate ``collapse''. Rather, the ``collapse'' is tied to the correlation between the system and the measurer -- to the recording by the observer of the measurement outcome. We note that -- from the point of view of the agent operating such an engine -- the collapse is irreversible as long as the observer retains the information about the outcome \cite{Zurek18}. This is in spite of the fact that the dynamics is unitary (hence, reversible), as can be at least in principle ascertained by the observer's friend who can reverse the measurement (providing he does not find out what was the outcome). 

One could equally well consider observers that measure coarse-grained energy of the gas particle rather than its coarse-grained location. One can also imagine an engine based on the energy measurement. Such a design would be closer to the original idea of Maxwell whose demon \cite{Maxwell} sorted hot from cold gas particles. 

This inability to ascribe a definite set of alternative states just on the basis of the density matrix alone is a reflection of the familiar {\it basis ambiguity} \cite{Zurek81}.  In our quantum engine, the density matrix will have the same eigenvalues of $\frac 1 2$ whether it is expressed in terms of the coarse-grained positions or suitably coarse-grained energy, so neither of these two alternatives can be considered preferred even if one ascribed (without justification!) special significance to the corresponding eigenstates.

Einselection of preferred states is the reflection of their ability to retain correlations with the apparatus or an observer in spite of decoherence \cite{Zurek81,Zurek82}. In the case of our quantum engine introduction of the partition in combination with the expectation that the interactions with the environment depend on distances will favor localized states, that is, $\bf L$ or $\bf R$. They will persist, as the tunneling between the two halves is suppressed by both the barrier and by the ongoing monitoring by the environment. Thus, after the barrier is introduced, the observer's ability to predict the future state of the gas particle (and, hence, the demon's ability to use measurement to extract work) will favor $\bf L$ vs $\bf R$ over coarse-grained measurement of energy. However, in the absence of the barrier, the coarse-grained measurement of energy may well be advantageous.

Our discussion also illustrates envariance for mixed states. The derivation of probabilities follows the same logic as for pure entangled states, although explicit entanglement is no longer essential -- suitable classical correlations are now enough. One can swap the location of the particle (e.g., ${\bf L} \rightleftharpoons {\bf R}$) and check if that swap can be undone by the corresponding counterswap on the bath. If the answer is ``yes'', the probabilities must be equal. We note, however, that in this case one starts the swap/counterswap sequence with a mixed state of the whole. Therefore, a confirmation that a swap was undone by a counterswap is no longer as straightforward as in the case of pure states. 

One can also take the view that all mixed states are -- in our quantum Universe -- mixed because of the entanglement with other systems. Indeed, this point of view (close to Ref. \cite{DeffnerZurek2016}) has been recently investigated in an effort to provide axiomatic foundations for statistical physics \cite{Chiribella}.

The quantum engine discussed here was recently employed (after suitable modifications) to help understand physical significance of quantum discord \cite{Zurekdiscord}, generalize Landauer's principle \cite{Ueda1}, investigate the role of indistinguishability in thermodynamic efficiency \cite{Ueda2}, study the role of quantum measurements \cite{Auffeves}, and investigate Gaussian entanglement \cite{Paternostro}. We have used it here to illustrate how statistical physics can arise in our quantum Universe without the artifice of ensembles. In retrospect, this ability to eliminate ensembles from quantum statistical physics is no surprise: quantum correlations are the origin of Born's rule and probabilities.  

We end by reiterating the distinction between the disappearance of the symptoms of quantumness (which are eradicated by decoherence) and the ``collapse'' that is responsible for the emergence of the definitive result. We note that -- in our Universe -- the same interactions that are responsible for decoherence usually disseminate, in many copies, the information about the preferred pointer observable of the system. This spreading of information was noted early on \cite{Zurek82} although its effects were more fully appreciated only more recently \cite{Zurek14,QD} and are still being investigated. In tandem with the realization that observers acquire their information about the world indirectly, such quantum Darwinism can explain all the symptoms of classicality, including the appearance of the ``collapse'' we perceive. 

Decoherence is an ongoing process that continues to extract the selected information about the system and deposit its copies in the environment, where they can be accessed by observers. If a demon attempted to measure an observable different from the pointer observable favored by decoherence, he would have to rely on direct measurements of the gas particle that would, moreover, lose their validity on the (presumably very short) decoherence timescale. Thus, only observables that commute with the pointer observable are worth observing, as only their records retain predictive power.

\ack{Enjoyable and informative discussions with Sebastian Deffner are gratefully acknowledged, as is the support by Department of Energy through the LDRD program at Los Alamos. This research was also supported, in part, by the Foundational Questions Institute grant number 2015-144057 on ``Physics of What Happens''.}

\section*{References}


\begin{thebibliography}{99}

\bibitem{Gibbs} J. W. Gibbs, {\it Elementary Principle is Statistical Physics} (Scribners, 1902).

\bibitem{Boltzmann} L. Boltzmann, Uber die Beziehung dem zweiten Haubtsatze der mechanischen W\"armetheorie und der Wahrscheinlichkeitsrechnung respektive den S\"atzen \"uber das W\"armegleichgewicht.
{\it Wiener Ber.} {\bf 76}, 373 (1877).

\bibitem{Maxwell} J. C. Maxwell, {\it Theory of Heat} (Longmans, 1871).

\bibitem{Zurek05} W.~H. Zurek, Probabilities from Entanglement, Born's Rule $p_k=|\psi_k|^2$
from Envariance. {\it Phys. Rev.} A {\bf 71}, 052105 (2005).

\bibitem{vonMises} R. von Mises, {\it Probability, Statistics, and Truth} (Dover, 1981).

\bibitem{Gnedenko} B.~V. Gndedenko, {\it The Theory of Probability} (Chelsea, New York, 1968).

\bibitem{Zurek03a} W.~H. Zurek, Environment-Assisted Invariance, Entanglement, and Probabilities in Quantum
Physics. {\it Phys. Rev. Lett.} {\bf 90}, 120404 (2003).

\bibitem{Zurek03b} W.~H. Zurek, 
Decoherence, Einselection, and the Quantum Origins of the Classical. 
{\it Rev. Mod. Phys.} {\bf 75}, 715 (2003).


\bibitem{Zurek11} W.~H. Zurek, Entanglement Symmetry, Amplitudes, and Probabilities: Inverting Born's Rule, {\it Phys. Rev. Lett.} {\bf 106} 250402 (2011).

\bibitem{DeffnerZurek2016} S. Deffner and W.~H. Zurek, Foundations of Statistical Mechanics from Symmetries of Entanglement, {\it N. J. Phys.} {\bf 18} (6) 063013 (2016).

\bibitem{Szilard} L. Szilard, 1929, On the Decrease of Entropy in a Thermodynamic System by the Intervention of Intelligent Beings, {\it Zeits. Phys.} {\bf 53} 840-856; English translation in {\it The Collected Works of Leo Szilard: Scientific Papers}, B.T. Feld and G. Weiss Szilard (eds.), pp. 103-129 (Cambridge, Massachusetts: MIT Press, 1972).

\bibitem{anders} S. Vinjanampathy and J. Anders, Quantum thermodynamics.
{\it Con. Phys.}  {\bf 57}(4)  pp. 545-579   (2016). 

\bibitem{ZP} W. H. Zurek and J.-P. Paz, Decoherence and the Second Law, {\it Phys. Rev. Lett.} {\bf 72}, 2508-2511 (1994).

\bibitem{Popescu} S. Popescu, A.J. Short and A. Winter, Entanglement and the Foundations of Statistical Mechanics {\it Nat. Phys.} {\bf 2}, pp. 754-758 (2006).

\bibitem{Goldstein} S. Goldstein, J.L. Lebowitz, R. Tumulka, and N. Zanghi, Canonical Typicality {\it Phys. Rev. Lett.} {\bf 96} 050403 (2006). 

\bibitem{Dirac} P. A. M. Dirac, {\it The Principles of Quantum Mechanics} (Clarendon Press, Oxford, 1976).

\bibitem{Bohr} N. Bohr, The quantum postulate and the recent development of atomic theory  {\it Nature} {\bf 121}, 580 (1928).

\bibitem{Everett} H. Everett III,
Relative State Formulation of Quantum Mechanics, 
{\it Rev. Mod. Phys.} {\bf 29}, 454 (1957).

\bibitem{Zurek81} W.~H. Zurek, Pointer Basis of Quantum Apparatus: Into What Mixture Does the Wave Packet Collapse? 
{\it Phys. Rev.} D {\bf 24}, 1516 (1981). 

\bibitem{Zurek91} W.~H. Zurek, Decoherence and the Transition from Quantum to Classical, {\it Physics Today}
{\bf 44}   Issue: 10   pp. 36-44   (1991).

\bibitem{GJK+96a}
E. Joos, H. D. Zeh, C. Kiefer, D. Giulini, J. Kupsch and I.-O. Stamatescu,
{\it Decoherence and the Appearance of a Classical World in
  Quantum Theory} (Springer-Verlag, Heidelberg, 2003).

\bibitem{Zurek14} W. H. Zurek, Quantum Darwinism, Classical Reality, and the Randomness of Quantum Jumps, {\it Physics Today} {\bf 47}, 44-50 (2014).

\bibitem{Zurek07} W.~H. Zurek,  Quantum Origin of Quantum Jumps: Breaking of Unitary Symmetry Induced by Information Transfer and the Transition from Quantum to Classical, {\it Phys. Rev.} A  {\bf 76}, 052110 (2007) 

\bibitem{Zurek13} W. H. Zurek, Wave-packet Collapse and the Core Quantum Postulates: Discreteness of Quantum Jumps from Unitarity, Repeatability, and Actionable Information, {\it Phys. Rev.} A {\bf 87}, 052111 (2013).

\bibitem{53} M. Schlosshauer, and A. Fine,  On Zurek's derivation of the Born rule, {\it Found. Phys.} {\bf 35}, 197-213 (2005).

\bibitem{5} H. Barnum, No-signalling-based Version of Zurek's Derivation
of Quantum Probabilities: A Note on ``Environment-Assisted Invariance,
Entanglement, and Probabilities in Quantum Physics'', quant-ph/0312150 (2003).

\bibitem{H} F. Herbut, Quantum Probability Law from 'Environment-Assisted Invariance' in Terms of Pure-State Twin Unitaries, {\it J. Phys.} A. {\bf 40}, 5949-5971 (2007).

\bibitem{Laflamme} L. Vermeyden, Xian Ma, J. Lavoie, M. Bonsma, Urbasi Sinha, R. Laflamme, and K.J. Resch, Experimental Test of Envariance, {\it Phys. Rev.} A {\bf 91}, 012120 (2015).

\bibitem{Ebrahim} J. Harris, F. Bouchard, E. Santamato, W. H. Zurek, R. W. Boyd, E. Karimi, Quantum Probabilities from Quantum Entanglement: Experimentally Unpacking the Born Rule, {\it N. J. Phys.} {\bf 18}, 053013 (2016).

\bibitem{Deffner} S. Deffner, Demonstration of Entanglement Assisted Invariance on IBM's Quantum Experience,
 {\it Helyion} {\bf 3}, e00444 (2017).

\bibitem{Ferrari} D. Ferrari, M. Amoretti, Demonstration of Envariance and Parity Learning on the IBM 16 Qubit Processor, arXiv:1801.02363 (2018).


\bibitem{Zurek82} W.~H. Zurek, Environment-induced Superselection Rules, 
{\it Phys. Rev.} D {\bf 26}, 1862 (1982).

\bibitem{Gleason} A.~M. Gleason, Measures on Closed Subspaces of Hilbert Space {\it J. Math. Mech.} {\bf 6}, 855-893 (1957).


\bibitem{40} P. S. Laplace, 1820, {\it A Philosophical Essay on Probabilities}, English translation of
the French original by F. W. Truscott and F. L. Emory (Dover, New York 1951).

\bibitem{Weinberg} S. Weinberg, {\it Lectures in Quantum Mechanics}, (Cambridge University Press, 2013).

\bibitem{HOM} C. K. Hong, Z. Y. Ou, and L. Mandel, Measurement of Subpicosecond Time Intervals Between Two Photons by Interference, {\it Phys. Rev. Lett.} {\bf 59} (18): 2044-2046 (1987). 

\bibitem{sample} For a sample of papers that make use of or comment  on the envariant derivation of Born's rule (not always agreeing with the views of this author) see: C. T. Sebens and S. M. Carroll, Self-locating Uncertainty and the Origin of Probability in Everettian Quantum Mechanics, {\it Brit. J. Phil. Sci.} {\bf 69} (1) 25-74 (2018); W. Son, Consistent Theory for Causal Non-locality Beyond the Born's Rule, {\it J. Korean Phys. Soc.}  {\bf 64}  (4) 499-503 (2014); C. Kiefer, Emergence of a classical Universe from quantum gravity and cosmology, {\it Phil. Trans. Roy. Soc.} A  {\bf 370}  (1975)  4566-4575  (2012); C. L. Hasse, Limits on the observable dynamics of mixed states, {\it Phys. Rev.} A {\bf 85} (6) 062124 (2012); S. D. H. Hsu, On the Origin of Probability in Quantum Mechannics, {\it Mod. Phys. Lett} {\bf 27} (12) 1230014 (2012); S. Olivares and M. G. A Paris, Entanglement-induced Invariance in Bilinear Interactions, {\it Phys. Rev.} A 
{\bf 80}  (3)  032329  (2009); A. M. Steane, Context, Spacetime Loops and the Interpretation of Quantum Mechanics, {\it J. Phys.} A {\bf 40} (12) 3223-3243 (2007); T. F. Jordan, Assumptions that Imply quantum dynamics is linear, {\it Phys. Rev.} A {\bf 73} (2) 022101 (2006).

 \bibitem{Schlosshauer} 
M. Schlosshauer, {\it Decoherence and the Quantum to Classical Transition}, (Springer, Berlin, 2007).

\bibitem{Toda} M. Toda, R. Kubo, and N. Saitöo, {\it Statistical Physics I}. (Springer, Berlin, 1983).

\bibitem{Umezawa} H. Umezawa, H. Matsumoto, and M. Tachiki, {\it Thermofield Dynamics and Condensate States} (North Holland, 1982).

\bibitem{Shannon} C. Shannon and W. Weaver {\it The Mathematical Theory of Communication}. (Urbana, Illinois: University of Illinois Press, 1949).

\bibitem{Landauer} R. Landauer,  
Irreversibility and Heat Generation in the Computing Process,
{\it IBM Journal of Research and Development} {\bf 5} (3), 183-191(1961); {\it Physics Today} {\bf 44} pp. 23-29 (1991).

\bibitem{Bennett} C. H. Bennett, 
Notes on the History of Reversible Computation,
{\it IBM Journal of Research and Development}, {\bf 17}, (6) 525-532 (1973); {\it ibid.} {\bf 32} (1), 16-23 (1988).

\bibitem{JauchBaron} J. M. Jauch and J. G. Baron, Entropy, Information, and the Szilard's Paradox, {\it Helv. Phys. Acta} {\bf 45}, 220-232 (1972).

\bibitem{ZurekSzilard} W. H. Zurek, Maxwell's Demon, Szilard's Engine and Quantum Measurements, pp. 151-161 in {\it Frontiers of Nonequilibrium Statistical Physics}, G. T. Moore and M. O. Scully, eds., (Plenum, 1986).

\bibitem{Koreans} Seung Ki Baek, Su Do Yi, and Minjae Kim, Particle in a Box with a Time-Dependent $\delta$-function Potential, 
{\it Phys. Rev.} A 94, 052124 (2016).

\bibitem{vonN} J. von Neumann,  
{\it Mathematical Foundations of Quantum Theory}, 
translated from German original by R. T. Beyer 
(Princeton University Press, Princeton, 1955).


\bibitem{Zurek89} W. H. Zurek, Algorithmic Randomness and Physical Entropy, {\it Phys. Rev.} A {\bf 40}, 4731 (1989).

\bibitem{d'Espagnat} B. d'Espagnat, {\it Veiled Reality} (WestviewPress, 2003).

\bibitem{ZP2} W.~H. Zurek and J.-P. Paz, Why We Don't Need Quantum Planetary Dynamics: Decoherence and the Correspondence Principle for Chaotic Systems, {\it Physica D} {\bf 83}, 300 (2015).

\bibitem{HuPazZhang} B.-L. Hu, J.-P Paz, and Y. Zhang, {it Phys. Rev.} D {\bf 45}, 2843 (1992). 

\bibitem{Yang} I.-S. Yang, The entanglement Timescale, arXiv:1707.05792 (2017).
 
\bibitem{ETH} There is not contradiction between the conservation of von Neumann entropy in closed quantum systems and the {\it eigenstate thermalization hypothesis} that examines properties of matrix elements of observable quantities in individual energy eigenstates of the system (see M. Srednicki, Chaos and Quantum Thermalization, {\it Phys. Rev.} E. {\bf 50} 888 (1994)). Indeed, (as it was recently suggested) one can formulate {\it eigenstate decoherence hypothesis} that parallels eigenstate thermalization hypothesis -- energy eigenstates of suitable fragments of a closed system anticipate the possibility of decoherence \cite{Lychkovskiy}.




\bibitem{Zurek18} W. H. Zurek, in preparation (2018).

\bibitem{Chiribella} G. Chiribella and C. M. Scandolo, Entanglement as an Axiomatic Foundation for Statistical Mechanics, arXiv:1608.04459

\bibitem{Zurekdiscord} J. Oppenheim, M. Horodecki, P. Horodecki, and R. Horodecki, A Thermodynamical Approach to Quantifying Quantum Correlations {\it Phys. Rev. Lett.} {\bf 89} (18), 180402 (2002); W.~H. Zurek, Quantum Discord and Maxwell's Demons, {\it Phys. Rev.} A {\bf 67}, 012320 (2003).

\bibitem{Ueda1} S. Kim, T. Sagawa, S. De Liberato, and M. Ueda,  
Quantum Szilard Engine,
{\it Phys. Rev. Lett.} {\bf 106}, 70401 (2011).

\bibitem{Ueda2} M. Plesch, O. Dahlsten, J. Goold and V. Vedral, Maxwell's Daemon: Information versus Particle Statistics, {\it Sci. Rep.} {\bf 4}, 6995 (2014); J. Bengtsson, M. N. Tengstrand, A. Wacker, P. Samuelsson, M, Ueda, H. Linke, and S. M. Reimann, Supremacy of the Quantum Many-body Szilard Engine with Attractive Bosons, arXiv:1701.08138 (2017).


\bibitem{Auffeves} N. Cottet, S. Jezouin, L. Bretheau, P. Campagne-Ibarcq, Q. Ficheux, J. Anders, A. Auff\'eves, R. Azouit, P. Rouchon, and B. Huard, 
    Observing a Quantum Maxwell Demon at Work, {\it PNAS} {\bf 114}, 7561 (2017).

\bibitem{Paternostro} M. Brunelli, M. G. Genoni, M. Barbieri, and M. Paternostro, Detecting Gaussian Entanglement via Extractable Work, {\it Phys. Rev.} A {\bf 96}, 062311 (2017). 

\bibitem{QD} 
H. Ollivier, D. Poulin, and W.~H. Zurek, 
Objective Properties from Subjective Quantum States: Environment as a Witness, 
{\it Phys. Rev. Lett.} {\bf 93}, 220401 (2004);  R. Blume-Kohout and W.~H. Zurek, 
Quantum Darwinism: Entanglement, branches, and the emergent classicality of redundantly stored quantum information, 
{\it Phys. Rev.} A {\bf 73}, 062310 (2006); J.~P. Paz and A.~J. Roncaglia, 
Redundancy of classical and quantum correlations during decoherence, 
{\it Phys. Rev.} A {\bf 80}, 042111 (2009); 
W.~H. Zurek, Quantum Darwinism, 
{\it Nature Physics}, {\bf 5}, 181 (2009); C. J. Riedel and W. H. Zurek, Quantum Darwinism in an Everyday Environment: Huge Redundancy in Scattered Photons, {\it Phys. Rev. Lett.} {\bf 105}, 020404 (2010); Redundant Information from Thermal Illumination: Quantum Darwinism in Scattered Photons, {\it N. J. Phys.} {\bf 13} 073038 (2011); M. Zwolak and W. H. Zurek  Complementarity of quantum discord and classically accessible information, {\it Sci. Rep.} {\bf 3}, 1729 (2013);
M. Zwolak, C. J. Riedel, and W. H. Zurek, Amplification, Redundancy, and the Quantum Chernoff Information, {\it Phys. Rev. Lett.} {\bf 112}, 140406 (2014); J. K. Korbicz, P. Horodecki, and R. Horodecki, Objectivity in the Photonic Environment Through State Information Broadcasting, {\it Phys.~Rev.~Lett.} {\bf 112}, 120402 (2014);
F.~G.~S.~L. Brandao, M. Piani, and P. Horodecki, Generic emergence of classical features in quantum Darwinism {\it Nat. Comm.} {\bf 6}, 7908 (2015); C. J. Riedel, W. H. Zurek, and M. Zwolak, The Objective Past of a Quantum Universe: Redundant Records of Consistent Histories, {\it Phys. Rev.} A {\bf 93}, 032126 (2016); C. J. Riedel, Classical branch structure from spatial redundancy in a many-body wavefunction, {\it Phys. Rev. Lett.} {\bf 118}, 120402 (2017);  G.  Pleasance  and  B.  M.  Garraway,  An  Application
of  Quantum  Darwinism  to  a  Structured  Environment, arXiv:1711.03732, (2017); P.  A.  Knott, T.  Tufarelli, M.  Piani, and G.  Adesso, Generic Emergence of Objectivity of Observables in Infinite Dimensions, arXiv:1802.05719 (2018).


\bibitem{Lychkovskiy} O. Lychkovskiy, Dependence of Decoherence-assisted Classicality on the Ways a System is Partitioned into Subsystems, {\it Phys. Rev.} A {\bf 87}, 022112 (2013); Decoherence at the Level of Eigenstates, arXiv:1712.04384 (2017).


   
%












\end{thebibliography}

\end{document}